\documentclass[floats,
            floatfix,
            showpacs,
            amssymb,
            prd,
            twocolumn,
            superscriptaddress,
            nofootinbib,
            nolongbibliography]{revtex4-2}

\usepackage{amssymb,
            amsmath,
            verbatim,
            mathtools,
            needspace,
            enumitem,
            etoolbox,
            graphicx,
            physics,
            microtype,
            afterpage,
            xspace,
            tabularx,
            lmodern,
            multirow,
            }
\usepackage[dvipsnames, usenames]{xcolor}

\definecolor{linkcolor}{rgb}{0.0,0.3,0.5}
\usepackage[unicode, colorlinks=true, linkcolor=linkcolor, citecolor=linkcolor, filecolor=linkcolor,urlcolor=linkcolor, pdfusetitle]{hyperref}
\usepackage[all]{hypcap}
\usepackage[T1]{fontenc}
\usepackage[utf8]{inputenc}
\usepackage[usenames,dvipsnames]{xcolor}

\usepackage{makecell}
\usepackage{mathrsfs}
\newcommand{\software}[1]{\texttt{#1}}

\definecolor{romared}{RGB}{142,0,28}
\hypersetup{colorlinks=true,
            citecolor=romared,
            linkcolor=romared,
            urlcolor=NavyBlue}

\newcommand{\nn}{\nonumber}

\newcommand{\be}{\begin{equation}}
\newcommand{\ee}{\end{equation}}

\def\be{\begin{equation}}
\def\ee{\end{equation}}
\newcommand{\beq}{\begin{eqnarray}}
\newcommand{\eeq}{\end{eqnarray}}

\newcommand{\dCS}{\mbox{\tiny dCS}}
\newcommand{\EdGB}{\mbox{\tiny EdGB}}
\newcommand{\PN}{\mbox{\tiny PN}}
\newcommand{\GR}{\mbox{\tiny GR}}
\newcolumntype{Y}{>{\centering\arraybackslash}X}

\newcommand{\icasu}{Illinois Center for Advanced Study of the Universe \&  Department of Physics, 
University of Illinois at Urbana-Champaign, Champaign, Illinois 61820 USA}
\newcommand{\xgi}{eXtreme Gravity Institute, Department of Physics, Montana State University, Bozeman, Montana 59717 USA}
\newcommand{\aei}{Max Planck Institute for Gravitational Physics (Albert Einstein Institute), Am M\"uhlenberg 1, 14476 Potsdam, Germany}
\begin{document}

\title{Improved gravitational-wave constraints on higher-order curvature theories of gravity}

\begin{abstract}
Gravitational wave observations of compact binaries allow us to test general relativity (and modifications thereof) in the strong and highly-dynamical field regime of gravity.
Here we confront two extensions to general relativity, dynamical Chern-Simons and Einstein-dilaton-Gauss-Bonnet theories, against the gravitational wave sources from the GWTC-1 and GWTC-2 catalogs by the LIGO-Virgo Collaboration.
By stacking the posterior of individual events, we strengthen the constraint on the square root of the coupling parameter in Einstein-dilaton-Gauss-Bonnet gravity to $\sqrt{\alpha_{\EdGB}} < 1.7$ km, but we are unable to place meaningful constraints on dynamical Chern-Simons gravity. 
Importantly, we also show that our bounds are robust to (i) the choice of general-relativity base waveform model, upon which we add modifications, (ii) unknown higher post-Newtonian order terms in the modifications to general relativity, (iii) the small-coupling approximation, and (iv) uncertainties on the nature of the constituent compact objects.
\end{abstract}

\author{Scott E. Perkins}
\email{scottep3@illinois.edu}
\affiliation{\icasu}

\author{Remya Nair}
\email{remya.nair@montana.edu}
\affiliation{\xgi}

\author{Hector O. Silva}
\email{hector.silva@aei.mpg.de}
\affiliation{\aei}
\affiliation{\icasu}

\author{Nicol\'as Yunes}
\email{nyunes@illinois.edu}
\affiliation{\icasu}

\date{\today}
\maketitle

%%%%%%%%%%%%%%%%%%%%%%%%%%%%%%%%%%%%%%%%%%%%%%%%%%%%%%%%%%%%%%%%%%%%%%%%%%%%%%%%%%%

%%%%%%%%%%%%%%%%%%%%%%%%%%%%%%%%%%%%%%%%%%%%%%%%%%%%%%
\section{Introduction}\label{sec:intro}
%%%%%%%%%%%%%%%%%%%%%%%%%%%%%%%%%%%%%%%%%%%%%%%%%%%%%%

General Relativity (GR) is our best description of gravity to date, fully consistent with all experiments performed to date. 
This is no small feat, as the precision and energy scales at which we now probe the gravitational interaction have been rapidly increasing. 
With the LIGO-Virgo collaboration (LVC) directly detecting gravitational waves (GWs)~\cite{LIGOScientific:2018mvr} through binary black hole (BBH), neutron star-black hole (NSBH), and neutron star-neutron star (NSNS) mergers, and the Event Horizon Telescope collaboration resolving the shadow of a supermassive black hole (BH)~\cite{Akiyama:2019cqa}, we are testing our best theory of gravity like never before.

With such impressive experimental support achieved by GR, one might wonder why resources should be expended on exploring alternative gravity theories.
From an observational standpoint, open questions like the existence or nature of dark matter~\cite{Sofue:2000jx,Bertone:2016nfn} and dark energy~\cite{Riess:1998cb,Perlmutter:1998np}, the quantum mechanical description of gravity, and the puzzle of the matter-antimatter asymmetry~\cite{Spergel:2003cb,Canetti:2012zc} are being widely studied.
For some of these questions, the problem might be resolved without additional forms of matter or energy, but rather by modifying the gravitational theory. 
This can come in the form of the addition of auxiliary fields and higher curvature terms to the gravitational action~\cite{Clifton:2011jh,Yunes:2013dva,Berti:2015itd}.
Beyond our desire to describe unexplained phenomena, there are purely null-test reasons to consider as well. 
The fundamental tenants of GR can be succinctly summarized by Lovelock's theorem~\cite{Lovelock:1971yv,Lovelock:1972vz}.
While GR is the only theory which satisfies the stipulations of this theorem, there is no fundamental reason why the true theory of gravity has to respect these criteria. 

A very interesting class of theories from these perspectives are known as quadratic theories of gravity~\cite{Yagi:2015oca}.
This class of theories introduce terms to the gravitational action that are quadratic in curvature invariants, composed of quantities like the Ricci scalar, Ricci tensor, and Riemann tensor.
Even though some of these theories lead to higher-than-second order field equations, these higher order terms can be tamed when treating them as arising from an effective field theory~\cite{Delsate:2014hba}. 
In this work, we focus on dynamical Chern-Simons (dCS)~\cite{Alexander:2009tp} and Einstein-dilaton-Gauss-Bonnet (EdGB)~\cite{Kanti:1995vq} theories, which have an extra scalar degree of freedom coupled to the Pontryagin density and Gauss-Bonnet density, respectively.

Modifications to the gravitational action, like those introduced in dCS and EdGB, will produce modified field equations and, in turn, result in dynamics that differ from those predicted in GR.
These difference can become imprinted in observables, like the GWs produced by inspiraling and merging binaries, which opens an avenue for discerning the true description of gravity obeyed by Nature~\cite{Yunes:2016jcc,Berti:2018cxi,Berti:2018vdi}.
Since LVC's first detection, the veritable treasure trove of information about strong field gravity contained in these signals has resulted in a gold rush of sorts, with a multitude of recent works looking for ``smoking-gun'' signatures of deviations from GR (see e.g.~\cite{Abbott:2020jks,TheLIGOScientific:2016src,TheLIGOScientific:2016pea,Abbott:2017vtc,Abbott:2017oio,Abbott:2018lct,Brito:2018rfr,Sennett:2019bpc,Yamada:2020zvt,Ghosh:2021mrv}), including our previous work specifically focused on the subject of quadratic theories of gravity~\cite{Nair:2019iur}.

\begin{table*}[ht]
    \centering
    \begin{tabular}{c  c  c  c}
        \hline \hline
        Theory & Coupling & New constraint & Prior constraint \\
        \hline
        EdGB & Gauss-Bonnet [Eq.~\eqref{eq:def_gaussbonnet}] & $1.7$ km & $2$--$5.6$ km~\cite{Nair:2019iur,Yagi:2012gp}\\ 
        dCS  & Pontryagin   [Eq.~\eqref{eq:def_pontryagin}]  & $-$      & 8.5 km~\cite{Silva:2020acr}   \\
         \hline\hline
    \end{tabular}
    \caption{ 
    Current state of quadratic theories of gravity and the bounds resulting from this work.
    The columns are, in order, the theory, the topological invariant that couples non-minimally to the scalar field, the constraints as the result of this work, and the previous strongest constraints on these theories.
    }
    \label{tab:summary}
\end{table*}

In this work, we first extend and confirm assertions made in our previous work~\cite{Nair:2019iur} by exploring the posterior surface of the parameter space through extensive numerical analysis of current LVC data and through a parametrization that is naturally tailored to this problem.
This removes any sampling artifacts that were introduced in our previous work as the result of recycling the theory agnostic analysis by the LVC.
With this in hand, we then place constraints on beyond-GR modifications through the combination of multiple detections using a fully Bayesian ``stacking'' of posterior probability density functions.
In all of these constraints, we ensure that our bounds are consistent with the small-coupling approximation, which is a necessary requirement for the waveforms used to be valid.
The key results of this analysis are summarized in Table~\ref{tab:summary}, which show that the stacked constraints are about 3 times stronger than previous GW constraints~\cite{Nair:2019iur} and about 15\% stronger than other model-dependent astrophysical bounds~\cite{Yagi:2012gp}. Therefore, the stacked constraints found in this paper are the strongest yet placed on this theory.

But are these constraints we find through stacking robust? We explore this question by studying the impact of unknown, higher-order post-Newtonian (PN) corrections introduced by beyond-GR modifications, as well as on systematic biases that could be introduced by using different GR base waveforms models.
The lack of information about higher PN order modifications has been of concern recently~\cite{Li:2011vx,Gupta:2020lxa,Julie:2019sab,Shiralilou:2020gah}. We show, however, that given basic assumptions about the nature of any mathematically consistent PN modification to the non-GR waveform, our results are robust. This means that, although the calculation of higher PN order terms in the non-GR waveform may be useful to better understand the late-inspiral and merger, these corrections are not necessary to place constraints on the theory because the constraints are dominated by the leading-order PN terms.  
%to any future work on finding higher order corrections with respect to the coupling constants.
%
Concerning the impact of the specific choice of waveform on our analysis, we compared the constraints obtained using two different GR waveform base models, onto which leading-order modifications are appended. We find that the impact of these systematics on the constraints we place are negligible. 
These results therefore prove that the stacked constraints we obtained here for EdGB gravity are robust to unknown PN corrections in both the GR and the non-GR sectors.

Can one place stronger constraints than these joint constraints by looking at particular events in the GWTC-2 catalog? Indeed, as pointed out in~\cite{Alexander:2017jmt}, there are some binary systems that, if detected, would be ideal to place constraints on quadratic gravity theories. In particular, the detection of a mixed (spinning BH and neutron star (NS)) binary could lead to much more stringent constraints than those obtained from BBHs alone. There are three events in the GWTC-2 catalog that could potentially be used for such tests: GW190425~\cite{Abbott:2020uma}, GW190426~\cite{Abbott:2020niy}, and GW190814~\cite{Abbott:2020khf}. Unfortunately, GW190425 is very likely to be a NS binary and not a mixed binary, in spite of the lack of an electromagnetic counterpart, because there are no known formation channels for BBHs in this mass range~\cite{Abbott:2020uma}. 
If GW190425 was produced by a NS binary, then monopole scalar charge would not be present~\cite{Yagi:2015oca,Wagle:2018tyk,Saffer:2019hqn}, thus preventing scalar dipole emission, and thereby suppressing non-GR modifications. Event GW190426 is probably a mixed binary, but unfortunately again, its false alarm rate (FAR) is extremely high ($1.4$~yr${}^{-1}$ - the highest rate in the entire catalog) and its signal-to-noise ratio (SNR) is not large enough ($10.1$)~\cite{Abbott:2020niy}, leading to very wide and somewhat uninformative posteriors on the extracted parameters.
In fact, the LVC analysis concerned with tests of GR \textit{excludes} this event due to its low significance. 
Finally, GW190814 can potentially be a BBH or NSBH binary, but the conclusions one arrives at differ drastically between the two assumptions.
Lacking a definite answer on the exact nature of this event, the results subsequently derived from this event are suspect.
We will show in this paper how all of this plays out (when considering EdGB theory in particular) and why these three events, among others, can therefore not be robustly used to place constraints on these theories. 

We emphasize that the prior assumptions on the composition of a binary can greatly affect the posterior conclusions obtained from parameter
estimation, and this fact is not new.
Indeed, imagine we would like to constrain the equation of state of supranuclear matter, instead of testing GR, and imagine we insisted on using GW190814 to do this. Through a Bayesian analysis, the LVC determined that this event was produced by an asymmetric binary with masses $\sim 23 M_\odot$ and $\sim 2.6 M_\odot$. The heavier object is obviously a BH. If we assume the lighter object is also a BH, then we cannot place any constraints on the NS equation of state. However, if we assume the lighter object is a NS, then we can place very stringent constraints on the equation of state, since many of them do not allow for NSs that massive. Clearly, given that the data cannot identify the nature of the object, we cannot use this event to constrain the equation of state. Constraining quadratic gravity with this event (and other similar ones) suffers from the same problems, and thus such constraints are not robust. 

The remainder of the paper is organized as follows.
In Sec.~\ref{sec:quad_grav}, we will outline the basics of the two theories we will be considering, including the GW phase modification introduced by them.
We will continue in Sec.~\ref{sec:methodology} with a discussion of the methodology (Sec.~\ref{sec:bayes_theorem}), preliminary estimates of our constraints (Sec.~\ref{sec:fisher}), our considerations when selecting sources (Sec.~\ref{sec:source_selection} and Appendix~\ref{app:source_classification}), and our final constraints (Sec.~\ref{sec:single_events} and~\ref{sec:multiple_events}) on the these theories.
Moreover, we establish the robustness of these results by analyzing the impact of waveform systematics and truncated expansions are discussed in Sec.~\ref{sec:gr_template} and Sec.~\ref{sec:GHO}, respectively.
Finally, we present our conclusions in Sec.~\ref{sec:conclusions}.
Throughout this work, we use geometrical units $G=c=1$.

%%%%%%%%%%%%%%%%%%%%%%%%%%%%%%%%%%%%%%%%%%%%%%%%%%%%%%
\section{Basics of Quadratic Gravity}\label{sec:quad_grav}
%%%%%%%%%%%%%%%%%%%%%%%%%%%%%%%%%%%%%%%%%%%%%%%%%%%%%%

We consider two higher-curvature modified theories of gravity, namely EdGB and dCS, which
are described by the action
\begin{equation}
    S = \int d^4 x \sqrt{-g} \left[ \kappa R 
    - (1/2) (\nabla \vartheta)^2
    + \mathscr{L}_{\EdGB, \dCS} \right]\,,
\end{equation}
where $\kappa = 1/(16 \pi)$, $g$ is the determinant of the metric $g_{\mu\nu}$,
and $\vartheta$ is a (pseudo) scalar field that couples to the Gauss-Bonnet $\mathscr{G}$
(Pontryagin ${}^{\ast}RR$) curvature invariants as\footnote{Strictly speaking, the name ``Einstein-dilaton-Gauss-Bonnet'' 
gravity refers to an interaction Lagrangian $\propto \exp(\vartheta)\, \mathscr{G}$~(see e.g.,~\cite{Kanti:1995vq}), which 
for small scalar field amplitudes reduces to the linearized coupling used here and oftentimes called ``shift-symmetric''
scalar-Gauss-Bonnet gravity in the literature.}
\begin{subequations}
\begin{align}
    \mathscr{L}_{\EdGB} &= \alpha_{\EdGB} \,\vartheta{\EdGB}   \, \mathscr{G}\,,
    \label{eq:def_gaussbonnet}
    \\
    \mathscr{L}_{\dCS}  &= (\alpha_{\dCS}/4)\, \vartheta{\dCS} \, {}^{\ast}RR\,,
    \label{eq:def_pontryagin}
\end{align}
\end{subequations}
where
\begin{equation}
{}^{\ast}RR = R_{\nu\mu\rho\sigma} {}^{\ast}R^{\mu\nu\rho\sigma}\,,
\end{equation}
constructed from the Riemann tensor $R_{\mu\nu\rho\sigma}$ and its dual ${}^{\ast}R^{\mu\nu\rho\sigma} = (1/2) \epsilon^{\mu\nu\alpha\beta}{R_{\alpha\beta}}^{\rho\sigma}$, where $\epsilon_{\mu\nu\rho\sigma}$ is the totally antisymmetric 
Levi-Civita symbol,
and
\begin{equation}
\mathscr{G} = R^2 - 4 R_{\mu\nu}R^{\mu\nu} + R_{\mu\nu\rho\sigma}R^{\mu\nu\rho\sigma}\,,
\end{equation}
where $R$ and $R_{\mu\nu}$ are the Ricci scalar and tensor.
Finally, $\alpha_{\EdGB}$ and $\alpha_{\dCS}$ are coupling constants, both 
with units of $[{\rm{length}}]^{2}$. 
GR is recovered in the limit $\alpha_{\EdGB, \dCS} \to 0$.

To ensure the perturbative well-posedness of both theories, we work in 
the small-coupling approximation, in which modifications to GR are considered small.
To establish the small-coupling approximation, it is convenient to define the dimensionless perturbative parameter
\begin{equation}
\zeta_{\dCS,\EdGB} = \alpha^{2}_{\dCS,\EdGB} / (\kappa \, {\ell}^4)\,,
\label{eq:def_zeta}
\end{equation}
where ${\ell}$ is the typical mass scale of a system.
For the small-coupling approximation to be valid, we must have $\zeta_{\dCS,\EdGB}<1$. 
Here we set a rough threshold for the validity of the approximation by requiring that
\begin{equation}
{\alpha}_{\dCS, \EdGB}^{1/2} / m_s \lesssim 0.5\,,
\label{eq:cutoff}
\end{equation}
where $m_s$ is the smallest mass scale involved in the problem. 
Hereafter, we use $\alpha$ to denote both $\alpha_{\dCS}$
and $\alpha_{\EdGB}$, where the distinction between these two should be obvious by context.

In both the theories we consider here, BHs support a nontrivial scalar field which is dipolar in dCS (see e.g.~\cite{Yunes:2009hc,Konno:2009kg}) and monopolar in EdGB (see e.g.~\cite{Kanti:1995vq,Yunes:2011we,Prabhu:2018aun}).
This results in the emission of scalar quadrupole (in dCS) and scalar dipole (in EdGB) radiation during BBH inspirals.
This additional channel for binding energy loss modifies the GW phase, with leading order correction appearing at 2PN\footnote{In the PN formalism, quantities of interest such as the conserved energy, flux, etc., can be written as expansions in $v/c$, where $v$ is the characteristic speed of the binary and $c$ is the speed of light. ${\cal{O}}((v/c)^n)$ corrections counting from the Newtonian (leading order GR) term are referred to as $(n/2)$-PN order terms~\cite{Blanchet:2013haa,Damour:2016bks}.} (for dCS) and -1PN (for EdGB) order.
In dCS gravity, the scalar field also results in a quadrupolar correction to the binary BH spacetime, introducing 2PN corrections to the binding energy, which in turn affect the GW phase evolution at the same PN order.

To be more precise, if we write the GW signal in the Fourier domain as
\be \label{eq:FD_waveform}
\tilde{h}(f) = A(f) \exp[i \Psi(f)]\,,
\ee
both the theories modify the phase schematically as
\be
\Psi = \Psi_{\rm GR} + \beta u^{b}\,,
\label{eq:ppe_phase}
\ee
i.e., by an additional contribution relative to the GR phase $\Psi_{\rm GR}$, where $u = (\pi {\cal M} f)^{1/3}$.
Here, ${\cal M} = m \eta^{3/5}$ is the chirp mass, $\eta = m_1 m_2 / m^2$ is the symmetric
mass ratio, and $m = m_1 + m_2$ is the total mass.
Equation~\eqref{eq:ppe_phase} is cast in the form suggested in the parametrized post-Einsteinian (ppE)
formalism~\cite{Yunes:2009ke,Chatziioannou:2012rf,Tahura:2018zuq}, where $b$ corresponds to the PN order at which the leading order correction modifies the GR phase $\Psi_{\rm GR}$ and $\beta$ is a constant that controls the amplitude of this modification.
In dCS and EdGB gravity, the exponent $b$ takes the values 
\be
b_{\dCS} = -1,
\quad
\textrm{and}
\quad 
b_{\EdGB} = -7\,,
\ee
respectively.
The amplitude coefficients $\beta$ for these theories are~\cite{Yagi:2011xp,Yagi:2012vf}
\begin{align}\label{eq:dCS_beta}
% ------------
\beta_{\dCS} &= - \frac{5}{8192} \frac{\zeta_{\dCS}}{\eta^{14/5}}
\frac{(m_1 s^{\dCS}_2 - m_2 s^{\dCS}_1)^2}{m^2}
+ \frac{15075}{114688} \frac{\zeta_{\dCS}}{\eta^{14/5}}
\nn \\
&\quad 
\times \frac{1}{m^2} \left(
m_2^2 \chi_1^2 - \frac{350}{201} m_1 m_2 \chi_1 \chi_2 + m_1^2 \chi_2^2
\right)\,,
\\\label{eq:EdGB_beta}
% ------------
\beta_{\EdGB} &= - \frac{5}{7168} \frac{\zeta_{\EdGB}}{\eta^{18/5}}
\frac{(m_1^2 s^{\EdGB}_2 - m_2^2 s^{\EdGB}_1)^2}{m^4}\,,
% ------------
\end{align}
where 
\be\label{eq:parallel_spin}
\chi_i = (\vec{S}_{i} \cdot \hat{L}) / m_i^2\,,
\ee
is the dimensionless spin parameter, obtained from the projection of 
the dimensionless spin angular momentum vector $\vec{S}_i$ onto the 
direction of the orbital angular momentum $\hat{L}$.
Finally, $s_i$ are the spin and mass dependent 
BH scalar charges, valid to all orders in spin~\cite{Yagi:2012vf,Yunes:2016jcc,Berti:2018cxi},
\begin{align}
s_i^{\dCS} &= \frac{2 + 2\chi_i^4 - 2 \varsigma_i - \chi_i^2 (3 - 2 \varsigma_i)}{2 \chi_i^3}\,,
\\
s_i^{\EdGB} &= \frac{2 \varsigma_i(1 - \varsigma_i) }{\chi_i^2}\,,
\end{align}
where $\varsigma_i = (1 - \chi_i^2)^{1/2}$.
The expression for $\beta_i^{\dCS}$ has uncontrolled remainders of 
${\cal O}(\chi^4)$ which have not been calculate yet, while the 
expression for $\beta_i^{\EdGB}$ is valid for all orders in spin.
We remark that the amplitude $A(f)$ is also modified in both theories, 
but these are subdominant relative to the changes to 
the phase~\cite{Tahura:2019dgr} and we do not consider them here.
Note that the EdGB and dCS modifications are completely controlled by the $\zeta$ parameter~\eqref{eq:def_zeta} (or equivalently by $\sqrt{\alpha}$), and thus, this is the only parameter that needs to be constrained by the data.

%%%%%%%%%%%%%%%%%%%%%%%%%%%%%%%%%%%%%%%%%%%%%%%%%%%%%%
\section{Constraints on Quadratic Gravity}\label{sec:methodology}
%%%%%%%%%%%%%%%%%%%%%%%%%%%%%%%%%%%%%%%%%%%%%%%%%%%%%%

\subsection{Bayesian inference in GW science}\label{sec:bayes_theorem}

Scientific conclusions drawn from GW observations are typically constructed through Bayesian inference. One can express Bayes' theorem in the context of GW analysis as:
\begin{equation}\label{eq:BT}
P(\boldsymbol{\theta}|D) = \frac{P(D|\boldsymbol{\theta}) P(\boldsymbol{\theta})}{P(D)}\,,
\end{equation}
where $D$ is the data coming from the detector network and $\boldsymbol{\theta}$ is the vector of parameters uniquely defining the waveform.
For this work, we use the \software{IMRPhenomPv2}~\cite{Hannam:2013oca,Khan:2015jqa,Husa:2015iqa} waveform, defined in GR by the parameter vector $\boldsymbol{\theta} =\lbrace \alpha', \delta, \psi, \iota, \phi_{\text{ref}}, t_{\text{c}},D_L, \mathcal{M}, \eta, a_1,a_2, \cos \theta_1, \cos \theta_2,\phi_1, \phi_2 \rbrace$, where $\alpha'$ and $\delta$ define the right ascension and declination of the binary in the sky (not to be confused with the $\sqrt{\alpha}$ coupling parameter of the modified theory), $\psi$ is the polarization angle defined with respect to the Earth centered coordinates (consistent with the definition in~\cite{ Anderson:2000yy}), $\iota$ is the inclination angle of the binary's orbital angular momentum at a reference frequency of $20$~Hz, $\phi_{\text{ref}}$ is the phase at a reference frequency, $t_{\text{c}}$ is the time of coalescence, $D_L$ is the luminosity distance, $\mathcal{M}$ is the detector frame chirp mass, $\eta$ is the symmetric mass ratio, $a_1$ ($a_2$) is the dimensionless spin magnitude of the larger (smaller) BH, $\theta_1$ ($\theta_2$) is the angle between the BH spin vector and the orbital angular momentum of the binary for the larger (smaller) BH, and $\phi_1$ ($\phi_2$) is the in plane orientation of the spin vector for the larger (smaller) BH.
To consider dCS (EdGB) waveforms, we append one additional parameter, $\sqrt{\alpha_{\dCS}}$ ($\sqrt{\alpha_{\EdGB}}$), which has units of length and defines the typical length scale at which these modifications take effect.

The terms in Eq.~\eqref{eq:BT} are identified as follows - $P(D|\boldsymbol{\theta})$ is the ``likelihood'' and describes the probability of observing the data $D$ given the parameters $\boldsymbol{\theta}$ of an asssumed model. 
In GW studies, this quantity is calculated for a given waveform model describing the GW, and a noise model that describes the detector characteristics.
$P(\boldsymbol{\theta}|D)$ is the ``posterior'' which is the quantity of interest in all parameter estimation studies. 
It is the probability density on the parameters $\boldsymbol{\theta}$ given the data $D$.
$P(\boldsymbol{\theta})$ is the ``prior'' which represents our knowledge of the parameters $\boldsymbol{\theta}$ before we analyze the data. 
This knowledge can be a result of a previous independent measurement or could be because of some theoretical/physical restrictions. 
For example, even without looking at the data, we can assume confidently that the mass of a BH is non-negative. 
For this work, we will use priors that are uniform in $\alpha'$, $\sin \delta$, $\phi_{\text{ref}}$, $t_c$, $\psi$, $\cos \iota$, $a_i$, $\cos \theta_i$, $\phi_i$, the component masses, $m_1$ and $m_2$, and $\sqrt{\alpha }$.
The prior on the luminosity distance is uniform in volume ($\propto D_L^2$).
Finally, $P(D) = \int P(D|\boldsymbol{\theta}) P(\boldsymbol{\theta}) d\boldsymbol{\theta} $ is the Bayesian evidence, which is the probability of obtaining the data $D$, marginalized over all possible parameter values for $\boldsymbol{\theta}$, and serves as the normalization constant. 

GW signals from the inspiral of compact binaries are extremely weak and the technique of matched filtering is used to extract these signals from the noisy data. 
Assuming Gaussian and stationary detector noise, this technique equates to defining the (log) likelihood function in Eq.~\eqref{eq:BT} as the following
\be\label{eq:likelihood}
\ln p(D|\boldsymbol{\theta}) \propto - (1/2) \sum_i^n \left( D_i-h | D_i-h\right)\,,
\ee
where $h$ is the template response function defined in Eq.~\eqref{eq:FD_waveform} contracted onto the $i$-th detector response matrix, and $n$ is the number of detectors in the network.
The noise weighted inner product is defined as 
\be\label{eq:inner_prod}
\left(A|B\right) = 4 \Re\left[ \int_0^\infty \frac{A B^{\ast}}{S(f)}\,df \right]\,,
\ee
where ${}^{\ast}$ denotes complex conjugation and $S(f)$ represents the power spectral density (PSD) function of the detector.

So far, we have focused on inferring the properties of a single binary system, but how can we combine the information from multiple observations to enhance our constraints?
A growing number of detections are joining the existing catalogs of GW sources available from the LVC.
The statistical power associated with a larger sample size can have a significant impact on the constraints on modified theories of gravity~\cite{Perkins:2020tra}.
This power can be harnessed through various methods that combine the statistical analysis of single events to produce tighter cumulative bounds. 
These methods vary depending on the type of parameter being constrained. 
For example, recent work has been done on a hierarchical framework to update constraints on parameters that are \emph{not} common between all the events being analyzed. 
The values these parameters take are then determined by their respective hyper distributions, where one would not expect each parameter to be the same between each source~\cite{Isi:2019asy}. 
This approach is suitable for the combination of information about generic deviations, for example parameters introduced by the ppE framework or the generic deviations investigated by the LVC~\cite{Abbott:2020jks}. 
We will focus on a simpler case, in which the parameter being constrained has a single, true value in Nature common to all events.
Both EdGB and dCS fall into this category, where deviations in the waveform are parametrized by a fixed value of the coupling constant $\sqrt{\alpha}$.

Consider a catalog of $N$ detections comprised of data $\{D_i\}$ described by parameters $\{\boldsymbol{\theta}_i\}$, where $i$ runs from 1 to $N$. 
The joint posterior for the parameters describing all the events can be written as 
\begin{equation}
    p(\sqrt{\alpha},\{\boldsymbol{\theta}_i\}|\{D_i\} ) = \frac{p(\sqrt{\alpha},\{\boldsymbol{\theta}_i\}) p(\{D_i\}|\sqrt{\alpha},\{\boldsymbol{\theta}_i\})}{p(\{D_i\})}\,,
\end{equation}
where we have explicitly separated the parameter of interest $\sqrt{\alpha}$, the only common parameter in the set, from the $N$ sets of source parameters $\{\boldsymbol{\theta}_i\}$.
To produce a cumulative bound on the parameter introduced by a modified theory of gravity, we need to marginalize over all parameters $\{\boldsymbol{\theta}_i\}$, which includes quantities like the binary masses, spins, sky location, and orientation for each binary in the catalog.
This leads to an intermediate expression for the marginal posterior distribution on $\sqrt{\alpha}$ given by
\begin{align}
    p(\sqrt{\alpha}|\{D_i\} ) &= \frac{p(\sqrt{\alpha}) }{p(\{D_i\})}\int p(\{\boldsymbol{\theta}_i\}) \nonumber \\
    &\quad \times p(\{D_i\}|\sqrt{\alpha},\{\boldsymbol{\theta}_i\}) d\{\boldsymbol{\theta}_i\} \,,
\end{align}
where we have assumed that the prior on the modified theory parameter is independent of the source parameters for all the observations and we note that the evidence, defined by the integral $p(\{D_i\}) = \int p(\{\boldsymbol{\theta}_i\},\sqrt{\alpha}) p(\{D_i\} | \{\boldsymbol{\theta}_i)\} d\{\boldsymbol{\theta}_i\} d\sqrt{\alpha}$, is just a normalizing constant that can be factored out of the integral. 

Because each of the events are statistically independent, we can separate the integral by source, such that 
\begin{align}\label{eq:final_alpha_posterior}\nonumber
    p(\sqrt{\alpha}|\{D_i\} )&= \frac{p(\sqrt{\alpha})}{p(\{D_i\})}
    \prod_i \int p(\boldsymbol{\theta}_i)p(D_i|\sqrt{\alpha},\boldsymbol{\theta}_i) d\boldsymbol{\theta}_i\,, \\
    &= \frac{p(\sqrt{\alpha})}{p(\{D_i\})}
    \prod_i p(D_i| \sqrt{\alpha})\,,
\end{align}
where $p(D_i|\sqrt{\alpha})$ is the partially-marginalized likelihood for the $i$-th event, marginalized over source parameters only.

%---------------------------------------------------------
\subsection{Selection of Sources}\label{sec:source_selection}

Our choice of events used in the Bayesian analysis carried out in this paper was guided by several motivations. 
Namely, we are confident that low mass events will typically be the most effective at constraining quadratic theories of gravity like dCS and EdGB, as the curvature is  higher for lighter systems. 
This idea is motivated by the form of the modifications shown in Eqs.~\eqref{eq:dCS_beta} and~\eqref{eq:EdGB_beta}. 
As the phase modification scales inversely with the total mass to the fourth power, smaller masses will drastically improve the constraint (all else being equal). 
This is still true when also considering the PN order of the modification, as the $\beta_{\dCS,\EdGB}$ phase factors in Eq.~\eqref{eq:dCS_beta} and Eq.~\eqref{eq:EdGB_beta} are accompanied by powers of $u=(\pi \mathcal{M} f)^{1/3}$. 
The phase modification (relative to the Newtonian term in GR proportional to $u^{-5}$) becomes $\beta_{\dCS} u^{4} $ and $\beta_{\EdGB} u^{-2}$ in dCS (2PN) and EdGB (-1PN), respectively.
In the case of EdGB, both components [the 
factor of $u$ and the specific form of Eq.~\eqref{eq:EdGB_beta}] improve the constraining power of lighter systems over heavier ones.
This is less obvious for dCS, where the modification enters at 2PN and introduces an additional factor of $u^{4} = (\pi \mathcal{M} f)^{4/3}$ relative to the Newtonian order term in GR.
This factor weakens constraints coming from light systems as compared with heavier systems.
However, combining the dependence on $u$ with $\beta_{\dCS}$ yields an overall dependence on mass of $m^{-8/3}$.
This leads to lighter system still outperforming heavier ones, in spite of the high PN order in which the modification is introduced.
This analytic reasoning is also supported by the numerical explorations performed in Refs.~\cite{Alexander:2017jmt,Perkins:2020tra}.
Furthermore, highly spinning systems would produce tighter constraints than non-spinning binaries as 
the phase modification grows with higher spin systems in general. 
Hence systems with large effective aligned spins $\chi_{\text{eff}}$ were also preferred, where $\chi_{\text{eff}}$ is defined as 
\begin{equation}
    \chi_{\text{eff}} = (m_1 \chi_1 + m_2  \chi_2) / m\,.
\end{equation}

Given all of this, there are several interesting sources in GWTC-1 and GWTC-2, but not all of these could be used to place constraints on dCS and EdGB gravity. 
To begin with, the highly asymmetric sources GW190814~\cite{Abbott:2020khf} and GW190412~\cite{LIGOScientific:2020stg} could be of interest, but systematics due to waveform mismodeling and the uncertain nature of the lighter compact object in GW190814 are of concern. 
These concerns (specifically in the case of GW190814) are explored more thoroughly in Appendix~\ref{app:source_classification}.

Another interesting source to analyze would be one of the first NSBH binaries observed to date, namely GW190426. Its extremely high false alarm rate (FAR) and its small SNR, however, makes any conclusions we might obtain from this data questionable.
On top of the lack of confidence we have that the signal is of astrophysical origin at all, the broadening of the mass distributions due to new degeneracies with the coupling parameters makes the analysis more complicated.
The widening of the component mass distributions can be severe enough to bring into question the nature of the lighter object, again introducing complications to our analysis similar to those seen with GW190814 and discussed in Appendix~\ref{app:source_classification}.

Finally, the source GW190425~\cite{Abbott:2020uma} was also excluded. 
While the prospect that this binary is actually of mixed nature (formed by a BH and a NS) would be exciting, that possibility seems remote.
According to the analysis performed by the LVC~\cite{Abbott:2020uma}, no known formation channels would be able to produce a BH of such a small mass as that found for the larger component in this event.
As it seems much more likely that this source is actually a pair of NSs, and NSNS binaries do not source dipolar radiation~\cite{Yagi:2015oca,Wagle:2018tyk,Saffer:2019hqn}, we omitted this source as well.

\begin{figure}
    \centering
    \includegraphics[width=\columnwidth]{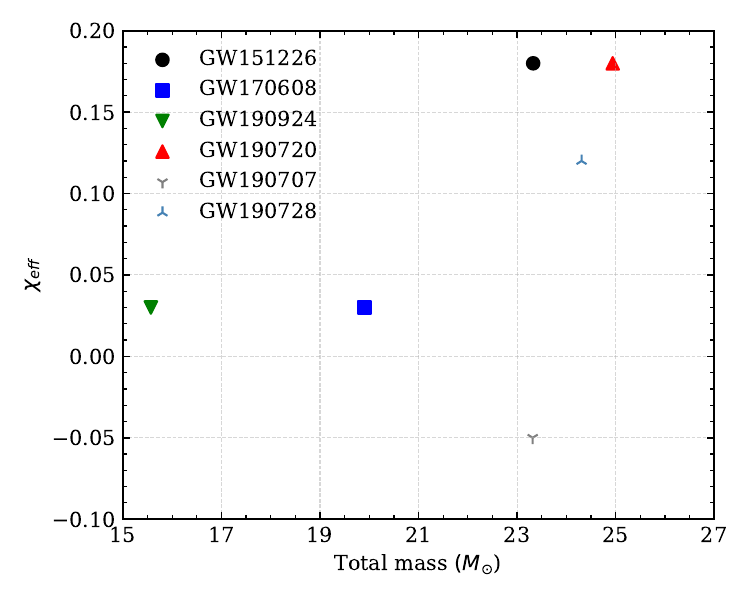}
    \caption{
    The detector frame total mass and the effective inspiral spin, $\chi_{\text{eff}}$, of the events analyzed in this work. 
    These values were inferred by LIGO assuming GR correctly describes the gravitational sector~\cite{Abbott:2020niy,LIGOScientific:2018mvr}.
    }
    \label{fig:source_properties}
\end{figure}

In summary, we have excluded the following events from our analysis: GW190814 and GW190412 (because their mass ratios are large, leading to issues with waveform systematics and complications to the analysis), GW190426 (because of a very high FAR, suggesting it may be a noise artifact) and GW190425 (because it is likely a NS binary and not a mixed binary). 
This then leaves the following set of sources to be analyzed in this work (we drop additional identifier information for GWTC-2 sources for brevity in the rest of this work): GW151226, GW170608, GW190924 (GW190924\_021846), GW190720 (GW190720\_000836), GW190707 (GW190707\_093326), and GW190728 (GW190728\_064510). 
Some of the critical source parameters discussed above are shown for each of the binaries analyzed in this study in Fig.~\ref{fig:source_properties}.

%---------------------------------------------------------
\subsection{Fisher analysis predictions}\label{sec:fisher}
It can be informative to determine an order of magnitude estimate of the constraints on $\sqrt{\alpha}$ before discussing (and performing) a full Bayesian analysis.
To this end, we use the more computationally tractable method of predicting the error covariances through the Fisher Information Matrix, or the Fisher for short~\cite{Cutler:1994ys,Finn:1992wt,Vallisneri:2007ev}.
This approximation begins by expanding the likelihood function about the maximum likelihood values $\boldsymbol{\theta}^{\text{ML}}$, resulting in the following expression
\be\label{eq:expanded_likelihood}
\ln p(D|\boldsymbol{\theta}) \approx -(1/2) \, \Gamma_{ij} \Delta \theta^i \Delta \theta^j\,,
\ee
where $\Delta \theta^i = \theta^{i,\text{ML}} - \theta^i$ is the deviation away from the maximum likelihood value, and $\Gamma_{ij}$ is the Fisher, 
\be
\Gamma_{ij} = \left( \partial_i h | \partial_j h \right)|_{\text{ML}} \,,
\ee
where $\partial_i = {\partial}/ {\partial \theta^i}$.
In the limit of large SNR and stationary Gaussian noise, the inverse of the Fisher gives the error covariance matrix $\Sigma$ of the parameters and the diagonal elements of this covariance matrix give the root mean square error in the estimate of the parameters:
\begin{equation}\label{eq:cov}
\sqrt{\left\langle(\Delta\theta^i)^2\right\rangle}=\sqrt{\Sigma^{ii}}\,.
\end{equation}

For the calculation of the Fisher, we use a restricted parameter set, mapping $\lbrace a_1, a_2, \cos \theta_1, \cos \theta_2, \phi_1, \phi_2\rbrace$ to $\lbrace \chi_1, \chi_2, \chi_\text{p}, \phi_\text{p}\rbrace$, where $\chi_i$ is the dimensionless spin defined in Eq.~\eqref{eq:parallel_spin}, $\chi_{\text{p}}$ is a certain projection of the BH spin onto the plane orthogonal to the binary's angular momentum, and $\phi_{\text{p}}$ is the angle of the perpendicular spin component $\chi_{\text{p}}$ in the plane.
The in-plane component of the spin is defined as in~\cite{LIGOScientific:2018mvr}, and can be written as 
\begin{equation}\label{eq:chip}
\chi_\text{p} = \frac{1}{B_1 m_1^2} \text{max}\left( B_1 S_{1 \perp}, B_2 S_{2 \perp} \right) \,,
\end{equation}
where $B_1 = 2 + 3 q / 2$, $B_2 = 2+ 3/ ( 2q )$, $q = m_2/m_1 < 1$ is the mass ratio and $S_{i\perp}$ is the projection of the spin of BH $i$ on the plane orthogonal to the orbital angular momentum $\mathbf{L}$.
We use a reduced parameter space because the Fisher can become ill-conditioned if the probability distributions have a high degree of covariance.  

The Fisher matrix must be evaluated at certain parameter values, called the injection parameters, which correspond to a rough guess of the maximum likelihood values of the posterior (around which the likelihood is expanded in the Fisher matrix approximation). 
Initially, the injection values for the Fisher matrices were taken from the LVC's published inferences on the values of the source parameters in order to quickly pick the most competitive sources.
However, in the course of validating our full analysis (detailed below), we derived posterior distributions for all the sources within GR.
As the LVC does not publish many of the less informative parameters, like the polarization angle, reference phase, etc., we recycled the posteriors from our custom GR analysis for the calculation of the Fishers presented in Table~\ref{tab:fisher} in order to have consistent injections.
The GR parameters were the median values recovered by our Markov-Chain Monte Carlo (MCMC) analysis, and the coupling parameter was set to zero, corresponding to our null hypothesis that GR is correct.

The noise curves used in the construction of the Fisher matrices were an O2 approximate PSD for Hanford and Livingston, and an O3-O4 proxy PSD for Virgo~\cite{ligo_SN_forecast}. 
The luminosity distances used in each Fisher were scaled such that the network SNR of the injection matched the SNR quoted by the LVC to account for minor differences in the detector noise.
To accurately model the network with which each source was observed, we calculate a separate Fisher matrix for each detector and combine them for each source before inverting the final matrix.
This gives a final Fisher for each source as 
\begin{equation}
    \Gamma_{\text{source}} = \sum_i^n \Gamma_{\text{source},i} \,,
\end{equation}
where $n$ is the number of detectors for each source, as reported by the LVC~\cite{Abbott:2020niy, LIGOScientific:2018mvr}.
Finally, the marginalized variance on the coupling constant from each of these covariance matrices, obtained from each source individually, are combined to construct a cumulative bound assuming a normal distribution. 
The process of inversion for the Fisher matrix plays the role of marginalization in this approximation, and the resulting components of the covariance matrix properly reflect uncertainties introduced through correlations. This gives the following relation:
\begin{equation}
\left(\, \Delta \sqrt{\alpha_{\text{total}}}\,\right)^{-2} 
=  \sum_i^N \left(\,\Delta \sqrt{\alpha_{i}} \,\right)^{-2}  \,,
\end{equation}
where $\Delta \sqrt{\alpha_{\text{total}}}$ is the final bound on the modifying parameter, $\Delta \sqrt{\alpha_{i}}$ is the standard deviation [defined by Eq.~\eqref{eq:cov}] for the $i$-th source, and $N$ represents the number of events.
The details of this calculation are outlined in Sec.~V and Appendix~A of~\cite{Perkins:2020tra}, and we refer the reader to that work for a more in-depth explanation of this analysis.

\begin{table}
\begin{tabular}{c| c c }
\hline \hline
Source & $(\Delta \sqrt{\alpha_{\EdGB}})$ km & $(\Delta \sqrt{\alpha_{\dCS}})$~km \\ \hline
GW151226 & 2.51 (6.02)  & 77.18 (6.02)\\
GW170608 & 2.98 (6.14)  & 41.38 (6.14)\\
GW190924 & 1.24 (4.02)  & 35.49 (4.02)\\
GW190720 & 3.30 (7.06)  & 35.53 (7.06)\\
GW190707 & 3.19 (6.83)  & 73.78 (6.83)\\
GW190728 & 3.17 (7.14)  & 69.0  (7.14)\\
\hline
Combined & 0.91 & 19.13 \\ \hline \hline
\end{tabular}
\caption{
Results from a Fisher projection, calculated using the median values for the source parameters from an MCMC analysis within GR. 
The results show the $90\%$ confidence interval for the root of the coupling parameter in km for each source.
The values in parenthesis are the upper limits on the validity of the small coupling approximation~\eqref{eq:cutoff}.
The last row of the table shows the combined constraint assuming each posterior is normally distributed, consistent with the assumptions of a Fisher analysis.
}\label{tab:fisher}
\end{table}

The Fisher results presented in Table~\ref{tab:fisher} suggest that stringent constraints on EdGB gravity may be possible for a variety of events. The same, however, is not true for dCS gravity, since the projected Fisher constraints all violate the small coupling approximation. Given this, we will next carry out a Bayesian analysis and compare our Bayesian constraints to the Fisher estimates of Table~\ref{tab:fisher}.

%---------------------------------------------------------
\subsection{Bayesian analysis: single events}\label{sec:single_events}

To sample from the posterior distribution for each source individually, we employed a MCMC analysis, which produces independent samples from the distribution of interest.
These samples can then be binned to show a discrete approximation to the posterior probability density.

\begin{figure}
    \includegraphics[width=\columnwidth]{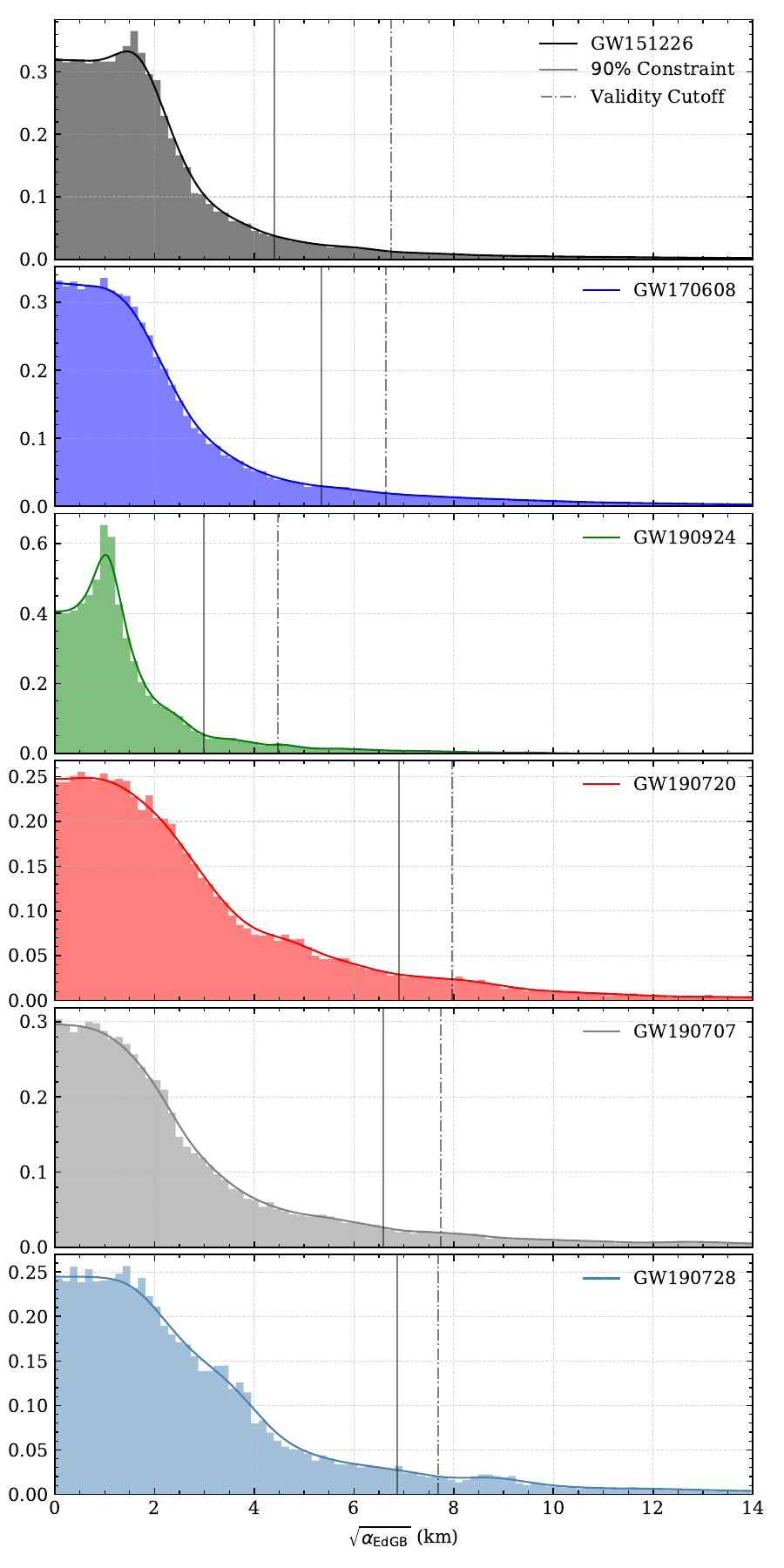}
    \caption{
    The individual constraints on the coupling constant of EdGB, $\sqrt{\alpha_{\EdGB}}$.
    Each panel shows the marginalized posterior distribution of the square root of the coupling constant in EdGB.
    Overlaid is the $90\%$ confidence value of the coupling constant, shown as the vertical solid line, with the upper limit of validity for the small coupling approximation shown as a vertical dashed line.
    On top of the discrete histogram, the KDE approximation used to determine the joint distribution from all the sources is also shown as a solid curve.
    All six sources shown satisfy the small coupling approximation at 90\% confidence, resulting in a robust bound on $\sqrt{\alpha_{\EdGB}}$.
}\label{fig:posterior_EdGB}
\end{figure}

\begin{figure}
    \includegraphics[width=\columnwidth]{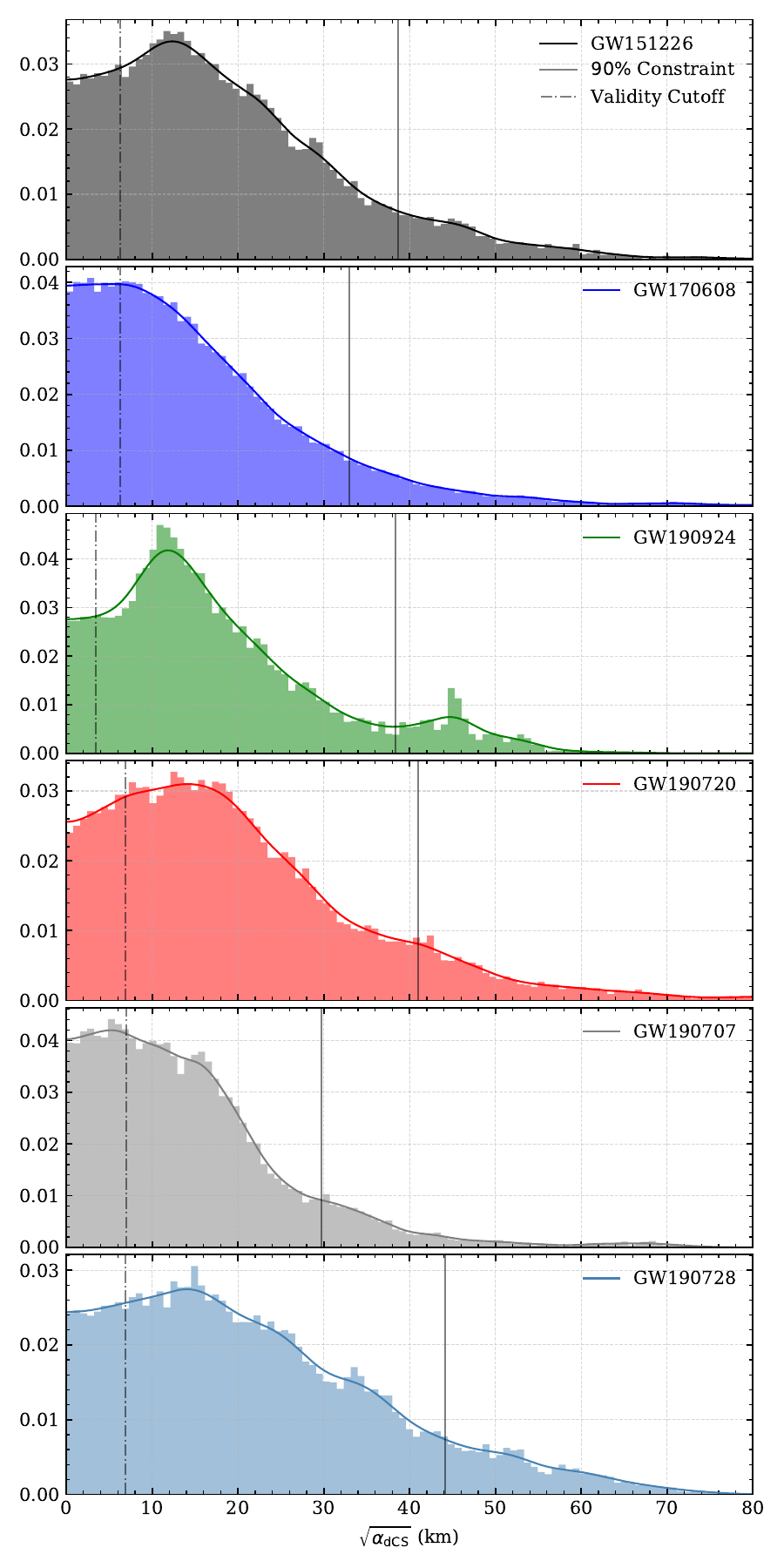}
    \caption{
    The individual constraints on the coupling constant of dCS, $\sqrt{\alpha_{\dCS}}$.
    Each panel shows the marginalized posterior distribution of the square root of the coupling constant in dCS.
    Overlaid is the $90\%$ confidence value of the coupling constant, shown as the vertical solid line, with the upper limit of validity for the small coupling approximation shown as a vertical dashed line.
    On top of the discrete histogram, the KDE approximation used to determine the joint distribution from all the sources is also shown as a solid curve.
    None of the sources shown satisfy the small coupling approximation, and as such, we can still not place a meaningful constraint on the dCS coupling parameter purely through GW observations.
    }\label{fig:posterior_dCS}
\end{figure}

The final one-dimensional marginalized posteriors on $\sqrt{\alpha}$ for EdGB (dCS) are shown in Fig.~\ref{fig:posterior_EdGB} (Fig.~\ref{fig:posterior_dCS}).
Each panel of the figures also shows the $90\%$ confidence upper limit on the magnitude of $\sqrt{\alpha}$, as well as the approximate upper limit on the validity of the small-coupling approximation in Eq.~\eqref{eq:cutoff}, required for the waveforms to be valid, both overlaid as vertical lines.
The approximate upper limit of the validity of the waveform was derived with the median value of the smaller mass from the same samples used to calculate the posterior distributions shown in Figs.~\ref{fig:posterior_EdGB} and~\ref{fig:posterior_dCS}.
We used as priors uniform on $\sqrt{\alpha}$ between $[0,20]$ km and $[0,100]$ km for EdGB and dCS, respectively. 
The fact that the posteriors in $\sqrt{\alpha_{\EdGB}}$ 
(i) differ considerably from the priors and 
(ii) the 90\% confidence upper limit is within the 
small-coupling approximation, give us confidence that these events can place a constraint on EdGB gravity.
The robustness of these constraints is analyzed in Sec.~\ref{sec:robustness}.

As can be seen from Fig.~\ref{fig:posterior_EdGB}, all the sources studied in this work independently place robust constraints on $\sqrt{\alpha_{\EdGB}}$, ranging between $3-7$ km.
The situation is opposite for dCS and none of the sources satisfy the small
coupling approximation.
As no individual analysis is trustworthy, any results derived from stacking these dCS posteriors would also not be correct. 
As such, the strongest constraint to date on this theory, remains the one established in~\cite{Silva:2020acr}, $\sqrt{\alpha_{\dCS}} < 8.5$~km, obtained by combining x-ray pulse profile observations of the milisecond pulsar PSR~J0030+0451 by NICER~\cite{Riley:2019yda,Miller:2019cac} with the NS tidal deformability inferred from the binary NS event GW170817 by the LVC~\cite{TheLIGOScientific:2017qsa,Abbott:2018exr}.
Any further discussion on the posteriors of $\sqrt{\alpha_{\dCS}}$ (combined or not) are pure speculation, for the benefit of future studies that might be more fortunate in the sources available for analysis.

Importantly, the sampling artifacts of the first iteration of this work~\cite{Nair:2019iur} have been removed through the MCMC analysis carried out here.
As speculated in~\cite{Nair:2019iur}, the lack of support at zero for GW151226 and GW170608 for both EdGB and dCS \emph{has disappeared with the use of the more natural sampling parameter $\sqrt{\alpha}$.}
Critically, the $90\%$ confidence regions of the different sources are consistent, indicating that, while the artifacts could be misleading, the conclusions reached in that work are substantiated. 

\subsection{Bayesian analysis: stacked events}\label{sec:multiple_events}

In theory, the process of combining posteriors from independent experiments to create a total probability distribution on a single parameter is as simple as multiplying the marginalized likelihoods together, considering Eq.~\eqref{eq:final_alpha_posterior}. 
To achieve this in practice two methods of combining posteriors are commonly employed, which we will briefly outline for clarity.

The first method begins by creating a histogram of the samples output by the MCMC analysis, which produces a discrete representation of the marginalized posterior on the parameter of interest.
One can then fit an ansatz function or use a more general kernel density estimation (KDE) to approximate the histogram, producing an analytic representation to the discrete posterior distribution.
This introduces two immediate sources of error, namely the possibility of a poor choice of fitting function and numerical errors in the fitting process directly.
For distributions that have obvious candidate ansatz (for example Gaussian distributions), this can be an attractive method as one can enforce certain properties, such as no-where vanishing probability distributions and smooth distributions for the final posterior.
When using KDE's, there is less flexibility to enforce these nice properties, like non-vanishing distributions or strict boundary conditions, but this method provides more flexibility in terms of minimizing manual ``tuning'' during the analysis.
This kind of technique is outlined and appropriately implemented in works like~\cite{DelPozzo:2011pg,Cardenas-Avendano:2019zxd}, for example.

A second method is even more straightforward, where the individual posteriors are also initially approximated by a histogram.
Instead of fitting a function to this discrete distribution, the histograms of each source are directly multiplied together (where, of course, the histograms must have the same bin sizes or be appropriately transformed). 
This method is convenient from the standpoint of how simple and easily implementable it can be.
However, numerical noise and finite numbers of samples can cause issues, especially when computing quantities at the tails of the distribution, like 95\% confidence intervals.
For example, if one source has zero counts in a certain histogram bin, the final distribution will \textit{always} have zero counts in the said bin, regardless of how much weight is assigned to that region by other sources.
This makes this particular method sensitive to finite bin-size and finite sample-size effects.
When calculating confidence intervals, smoothing functions can also be applied to the final distribution to minimize issues with convergence for numerical integration, which was a step implemented for this work.

\begin{figure}[t]
    \includegraphics[width=\columnwidth]{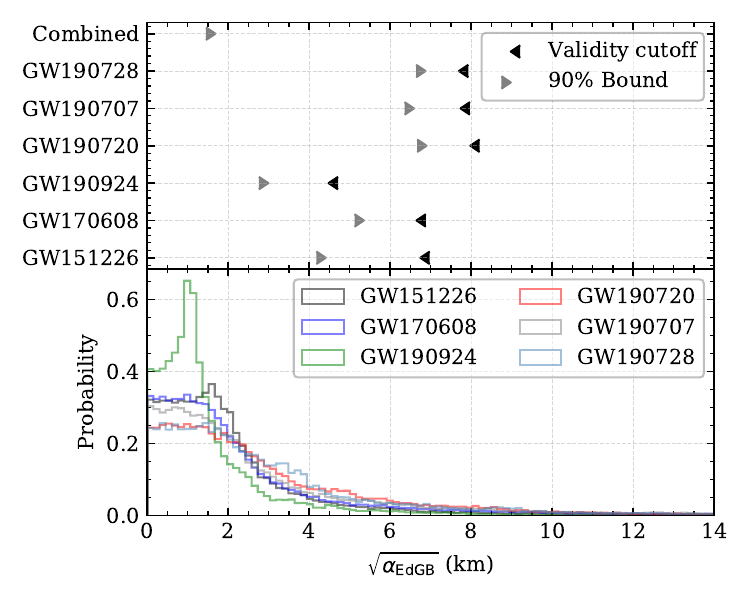}
    \caption{
    In the bottom panel, we show the histogram representation of the probability density of the value of the EdGB coupling constant, $\sqrt{\alpha_{\EdGB}}$, for the six events we have chosen for this analysis.
    The top panel shows the 90\% confidence constraint on the magnitude of the coupling parameter and the maximum value for validity of the small-coupling approximation for each of the individual events, with the cumulative constraint shown at the top.
    After combining the information of all the sources, we can achieve a constraint on $\sqrt{\alpha_{\EdGB}}$ of less than $1.7$ km at 90\% confidence.
    }\label{fig:combined_EdGB}
\end{figure}
\begin{figure}[t]
    \includegraphics[width=\columnwidth]{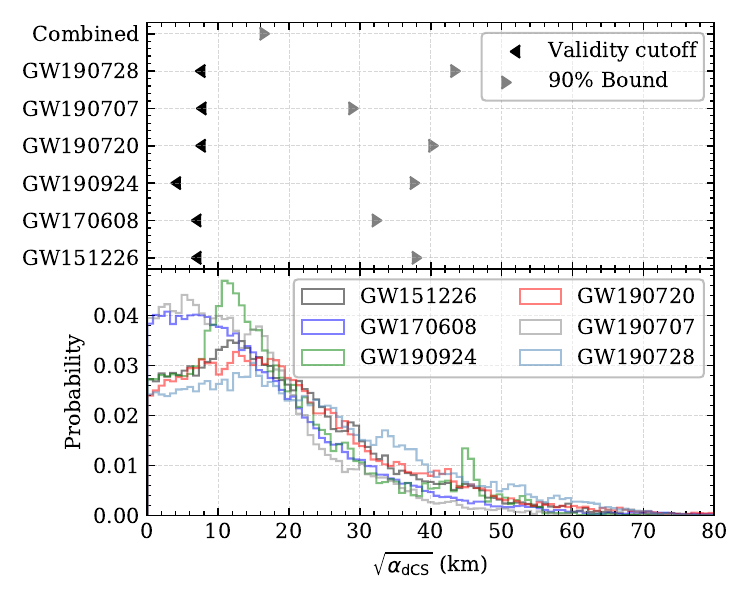}
    \caption{
    In the bottom panel, we show the histogram representation of the probability density of the value of the dCS coupling constant, $\sqrt{\alpha_{\dCS}}$, for the six events we have chosen for this analysis.
    The top panel shows the 90\% confidence constraint on the magnitude of the coupling parameter and the maximum value for validity of the small-coupling approximation for each of the individual events, with the cumulative constraint shown at the top.
    As no single constraint satisfies the small-coupling approximation, our cumulative bound is also untrustworthy. 
    We only show the combined posterior distribution for illustrative purpose.
    }\label{fig:combined_dCS}
\end{figure}

To ensure maximum reliability, we have repeated the calculation twice, once with each method, and always quote the more pessimistic constraint on $\sqrt{\alpha}$ throughout this work.
% For a discussion on the differences on these techniques in the context of this work, we refer the reader to Appendix~\ref{app:stacking_comparison}.
We discuss the difference of these techniques in the context of this work 
in Appendix~\ref{app:stacking_comparison}.

The results of combining posteriors on $\sqrt{\alpha}$ in EdGB and dCS are shown in Fig.~\ref{fig:combined_EdGB} and Fig.~\ref{fig:combined_dCS}, respectively.
We see a moderate improvement, as expected with a set of posteriors that are already comparable in constraining ability~\cite{Perkins:2020tra}.
Our new bound on the coupling parameter in EdGB at 90\% confidence is $\sqrt{\alpha_{\EdGB}} < 1.7$ km.
For dCS, as all the constituent posteriors in the set violate the small coupling approximation, \emph{the cumulative bound is still not valid and is only shown for illustrative purposes and guidance for future efforts to constrain this elusive theory with GW observations.}

%%%%%%%%%%%%%%%%%%%%%%%%%%%%%%%%%%%%%%%%%%%%%%%%%%%%%%
\section{Robustness of Constraints}
\label{sec:robustness}
%%%%%%%%%%%%%%%%%%%%%%%%%%%%%%%%%%%%%%%%%%%%%%%%%%%%%%

The constraints we have placed here are contingent on the reliability of each individual component in our analysis.
Undoubtedly, the GW waveforms we have used are one of the most uncertain elements.
There are two factors that could play a role in causing systematic biases in our conclusions. 
The first issue we address concerns the base model (that is, the waveform model in GR onto which our modifications are appended) implemented for this work, as there are many variations available today~\cite{Isoyama:2020lls}. 
This is discussed below in Sec.~\ref{sec:gr_template}. 
We then examine the issue of missing information at higher PN orders sourced by the contending theories, discussed below in Sec.~\ref{sec:GHO}.
In both sections, we will focus on the bounds we have placed upon the coupling constant in EdGB, as we are still unable to place constraints on dCS gravity with GW observations alone.

%---------------------------------------------------------
\subsection{Changing the GR base model}\label{sec:gr_template}
%---------------------------------------------------------

As mentioned earlier, we have used a widely adopted waveform as our base GR template, \software{IMRPhenomPv2}~\cite{Hannam:2013oca}. By base model, we mean that this is the model that we added the EdGB and dCS corrections to. However, there exist other templates that we could have used, such as some that include higher-order spin effects, higher harmonics, and various effective-one-body models.
The custom software developed for this analysis currently lacks support for most of these other options but does include the \software{IMRPhenomD}~\cite{Khan:2015jqa,Husa:2015iqa} template.
To quickly infer an estimate of the impact of the choice of base waveform template on our results with the tools available, we have rerun our analysis with \software{IMRPhenomD} as our base GR model.
The comparison between using an \software{IMRPhenomPv2} and \software{IMRPhenomD} base GR model will provide some insight into how sensitive our constraints are to the underlying GR base waveform.

The two base GR waveforms we will use, \software{IMRPhenomPv2} and \software{IMRPhenomD}, differ by the inclusion of precession effects. 
The \software{IMRPhenomD} waveform can only model spin-aligned and spin-anti-aligned binaries, whereas the \software{IMRPhenomPv2} waveform  can deal with precession effects by ``twisting up'' a co-precessing waveform (in this case \software{IMRPhenomD}) through frequency-dependent Euler angles~\cite{Hannam:2013oca}.
This could impact our analysis in one of the two ways: the extra information imparted by precessional effects could break certain degeneracies between source parameters, resulting in tighter bounds. 
Alternatively, it could be that the increased dimension of the model (moving from two spin parameters to six) could cause a degradation in accuracy of recovered parameters due to new degeneracies.

\begin{figure}
    \includegraphics[width=\columnwidth]{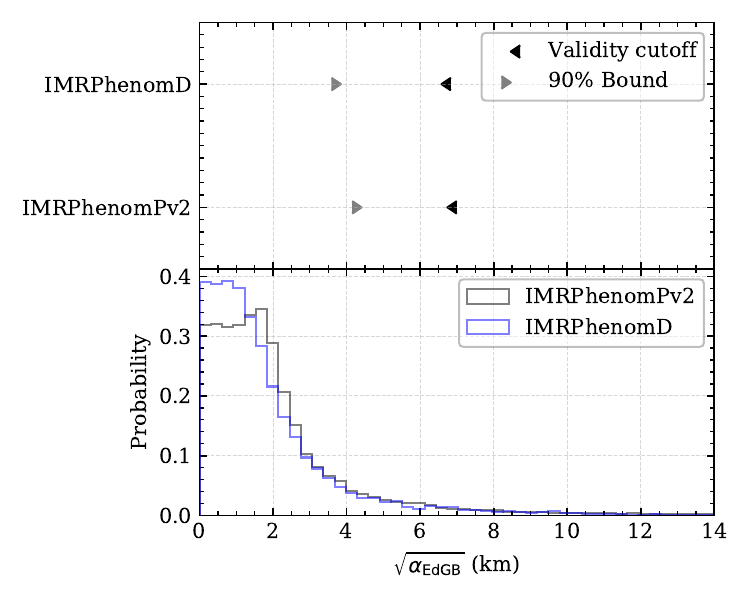}
    \caption{
    Comparison of the constraints obtained through two separate analyses of the event GW151226.
    In the first iteration, we used a more robust template, \software{IMRPhenomPv2}, which encodes information about precession.
    We then repeated the analysis using the base template \software{IMRPhenomD}, which only models spin-aligned and spin-anti-aligned binaries.
    Our results indicate that the constraints we have placed on the coupling constant of EdGB are robust to changes in the GR-template used to recover the parameters of the signal.
    }
    \label{fig:D_P_comparison}
\end{figure}

Our results are shown in Fig.~\ref{fig:D_P_comparison}, which presents the marginalized posteriors on $\sqrt{\alpha_{\EdGB}}$ using both an \software{IMRPhenomPv2} base model and a \software{IMRPhenomD} base model, and the GW151226 event as an example.
As the figure shows, the posteriors lead to very similar constraints, with the \software{IMRPhenomPv2} base model leading to $\sqrt{\alpha_{\EdGB}} \lesssim 4.5$ km and the  \software{IMRPhenomD} base model leading to $\sqrt{\alpha_{\EdGB}} \lesssim 3.5$ km. 
For this example, we see that the GW151226 event did not contain enough precessional information to lead to a better constraint; in fact, the increase in dimensionality of the parameter space worsens the constraint by about 20\%. 
Such a deterioration of the constraint is very small, and using the more complex model leads to a more conservative (i.e.~less stringent) constraint. 
This, therefore, increases our confidence that the constraints we have placed here are robust against the choice of base GR waveform model.

%---------------------------------------------------------
\subsection{Adding higher-order PN order GR modifications}\label{sec:GHO}
%---------------------------------------------------------
The waveforms we have used here only include the leading-order PN correction due to EdGB and dCS in the inspiral of the binary.
As there are currently no \textit{complete}, analytic expressions for higher-order corrections,
despite ongoing efforts both with PN methods~\cite{Yagi:2011xp,Loutrel:2018rxs,Loutrel:2018ydv,Julie:2019sab,Shiralilou:2020gah} and numerical relativity approaches~\cite{Witek:2018dmd,Okounkova:2019dfo,Okounkova:2019zjf,Okounkova:2020rqw,East:2020hgw,Silva:2020omi},
we are forced to work with a template which we know is lacking far behind the accuracy of the base (GR) template.

Given this uncertainty about the true waveform, there could be some concern that the inclusion of higher PN order corrections in the modifications to GR could degrade our bound through increased degeneracy between source parameters and the coupling parameter.
However, it is reasonable to expect that this will not be the case when considering realistic higher PN order corrections~\cite{Yunes:2016jcc}, as there are some fundamental restrictions that these additional modifications must satisfy to remain mathematically consistent in PN theory.

There are two criteria that any modification to GR that admits a PN expansion must meet: (i) all corrections to the waveform must be linear in the expansion parameter $\zeta$, and (ii) the expansion must be valid in the typical domain in which PN approximations are applied. The first condition must be true if one is working in the small coupling approximation. The second condition requires that the coefficient in the PN expansion of the modified gravity terms do not increase with PN order much more rapidly than how the coefficients grow in GR. Since these coefficients can grow in GR quite rapidly (especially in the extreme mass ratio limit), this second condition is not as stringent as it may seem at first.  

With this in mind, the additional terms entering the waveform would need to have a very particular and unfortunate functional form to \textit{increase} degeneracies, as opposed to decreasing them.
This is because the only way for additional information in the waveform (introduced through the higher PN order corrections) to increase the ambiguity about source parameters is to have very precise cancellation between terms. 
As this seems unlikely given the additional information is proportional to a different power of velocity (or frequency), when higher PN order corrections are finally  derived, they will only serve to improve our current constraints. Therefore, not including such higher PN order terms will lead to weaker or less stringent constraints, which we call here ``conservative.''

Lacking a complete higher PN order template to compare against, we will here approximate the degree of systematic bias we may be incurring by truncating the non-GR modification at leading PN order. 
We will do so by appending reasonable, generic modifications at higher PN orders as place holders for the true modifications.
These generic place holders can then be marginalized over, leaving a final cumulative bound on the coupling parameter that better reflects uncertainties in the higher PN order corrections omitted by our waveforms.

\begin{figure*}
    \centering
    \includegraphics[width=\textwidth]{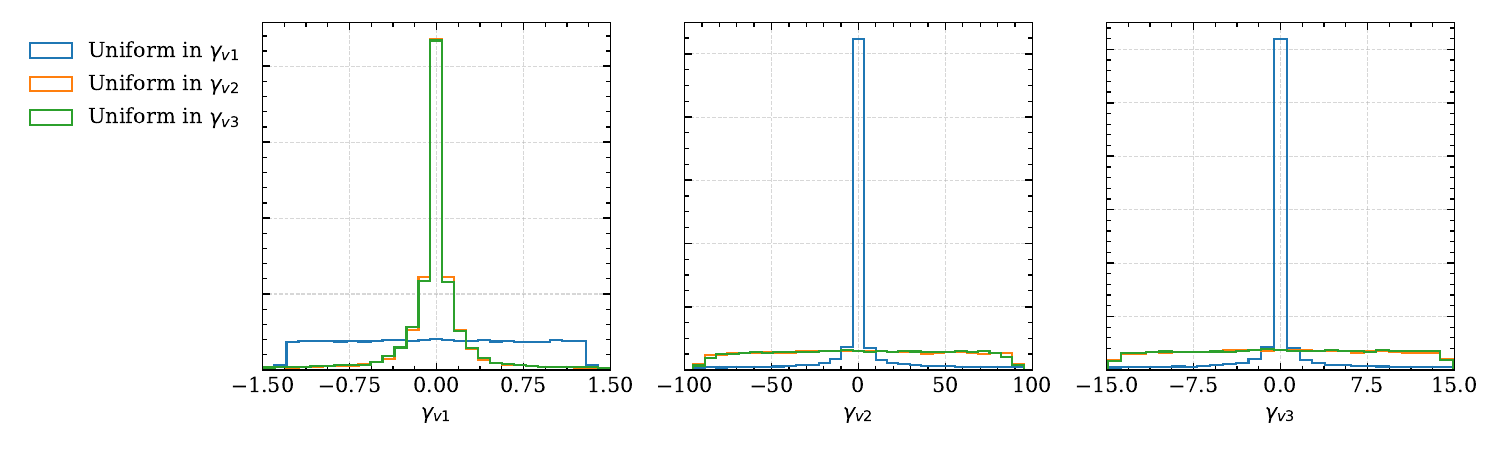}
    \caption{Comparison of priors for the different parametrization schemes outlined in Sec.~\ref{sec:GHO}. 
    The left panel shows the priors of each parametrization transformed to the parametrization $\delta\phi_{{\rm v}1}$, shown in Eq.~\eqref{eq:GHOv1}.
    The middle panel shows each parametrization transformed to that shown in Eq.~\eqref{eq:GHOv2}.
    Finally, the right panel shows the priors transformed to the parametrization shown in Eq.~\eqref{eq:GHOv3}.
    From these figures, we can see that the first parametrization differs substantially from parametrizations 2 and 3, while parametrizations 2 and 3 are very similar.
    }
    \label{fig:GHO_priors}
\end{figure*}

Marginalizing over the generic higher PN order term leads to a final constraint on the coupling parameter that reflects uncertainties in this new parameter, but even generic parametrizations that are marginalized over are not created equal. 
We have noted these differences in previous works~\cite{Nair:2019iur}, where it was shown that different generic parametrizations can result in different final constraints through the implicit difference in priors.
While this is not always a significant issue, as re-weighting the samples with new priors can sometimes mitigate this effect, our previous work showed a particularly unfortunate case where the Jacobian of the transformation from one parametrization to another was singular.

This is not something that can be unambiguously resolved, and so in this work we will pick three well-motivated parametrizations and compare the final results.
As outlined above, an important aspect that is shared by all three parametrizations is the fact that modifications at all orders should be proportional to the non-GR coupling parameter, in this case $\zeta_{\EdGB}$.
This must be the case for any mathematically consistent, higher PN order expansion of the EdGB field equations.

We will use the variable $\gamma$ to represent any higher PN order coupling of source parameters to the coupling parameter, which will be the variable that we will marginalize over by the end of this analysis.
This is a conservative implementation, as this parameter would be replaced by a function of the conventional source parameters in the true expansion, just as is the case at -1PN in the standard EdGB waveform.
As pointed out earlier, information introduced by this new coupling in the true expansion will almost certainly reduce our estimate of $\sqrt{\alpha_{\EdGB}}$, not widen it.
This ultimately can be seen from the fact that if higher PN order terms, proportional to the expansion parameter, are appended to the waveform, the waveform will deviate further from the GR template for equal values of the expansion parameter. 
This should only serve to reduce uncertainty in the modifying variable.

Beyond these details, it is also important to note that the true expansion (as of yet unknown) would \textit{not} increase the dimensionality of the parameter space. 
Any higher order corrections would only depend on the existing source parameters and the coupling constant, $\sqrt{\alpha_{\EdGB}}$.
Because of these considerations and the fact that our analysis here \textit{does} add an additional parameter, our results here can be viewed as even more conservative.

All three parametrizations versions can be written as in the form 
\begin{equation}
\delta \phi_{\text{v}i} = \phi_{\text{GW}} - \phi_{\GR}\,,
\quad (i = 1,2,3)\,.
\end{equation}
The first parametrization we will use modifies the GW phase as follows 
\begin{align}\label{eq:GHOv1}
    \delta \phi_{\text{v1}} 
    = \zeta_{\EdGB} \,  g_{\EdGB}(\boldsymbol{\theta}_{\GR}) \, u^{-7} + \zeta_{\EdGB} \, \gamma_\text{v1} \, u^{-5}\,,
\end{align}
where $u = (\pi \mathcal{M} f)^{1/3}$ as before 
and $g_{\EdGB}(\boldsymbol{\theta}_{\GR})= \beta_{\EdGB} / \zeta_{\EdGB}$ is the source dependent term in the leading order deviation due to EdGB defined in Eq.~\eqref{eq:EdGB_beta}.
This term depends on the parameters of the binary described within GR, $\boldsymbol{\theta}_{\GR}$. The second term in the above equation therefore introduces a 1PN correction to the -1PN dipole EdGB correction to the GR waveform phase, with $\gamma_\text{v1}$ assumed to be a constant number we will later marginalize over.
This choice is motivated by the ppE framework, where the additional modification can be seen as an absolute deviation.

The second parametrization we will investigate modifies the GW phase in the following way
\begin{align}\label{eq:GHOv2}\nonumber
    \delta \phi_{\text{v2}} &=\zeta_{\EdGB} \, g_{\EdGB}(\boldsymbol{\theta}_{\GR}) \, u^{-7}
    \\
        &\quad 
        + \zeta_{\EdGB} \, \gamma_\text{v2} \,  g_{\EdGB}(\boldsymbol{\theta}_{\GR}) \, u^{-5}\,.
\end{align}
Making the 0PN modification proportional to the same function as the leading PN order modification results in a relative deviation from the leading PN order term.
For clarity, we can rewrite the phase slightly to give the overall phase modification as 
\begin{equation}
\delta \phi_\text{v2} = \zeta_{\EdGB} g_{\EdGB}(\boldsymbol{\theta}_{\GR}) u^{-7} [ 1 + \gamma_\text{v2} u^{2}]\,,
\end{equation}
which shows more clearly the motivation of this parametrization. As before, here $\gamma_\text{v2}$ is also a constant we will later marginalize over. 

Finally, the third parametrization we will investigate modifies the GW phase as follows
\begin{align}\label{eq:GHOv3}\nonumber
    \delta \phi_{\text{v3}} &=  \zeta_{\EdGB} \, g_{\EdGB}(\boldsymbol{\theta}_{\GR}) \, u^{-7}\\
        &\quad+ \zeta_{\EdGB} \, \gamma_\text{v3} \, g_{\EdGB}(\boldsymbol{\theta}_{\GR}) \, \phi_{\text{GR,1\PN}} (\boldsymbol{\theta}_{\GR}) \, u^{-5}\,,
\end{align}
where the term $\phi_{\text{GR,1\PN}}$ is the 1PN term appearing in the PN expansion in GR, $\phi_{\text{GR,1\PN}}= 3715/756 + 55 \eta / 9$.
This choice in parametrization is motivated by the idea that the proper expansion of the EdGB field equations might mirror the form taken by the same expansion in GR, meaning the generic scheme imposed here would match the true expansion in EdGB when $\gamma_{\text{v3}} \rightarrow 1$.
This type of modification can be rewritten in the following form, 
\begin{equation}
\delta \phi_{\text{v3}} = \zeta_{\EdGB} g_{\EdGB}(\boldsymbol{\theta}_{\GR}) u^{-7}  [ 1 + \gamma_{\text{v3}} \phi_{\text{GR,1\PN}}u^{2}]\,,
\end{equation}
showing this relation more clearly, where once more $\gamma_\text{v3}$ is a constant we will later marginalize over.

As we stated above, we will marginalize over the $\gamma$ parameters, which control the size of the 1PN modifications considered here, relative to the leading PN order modification. But how large should we allow these $\gamma$ parameters to be? The answer to this question is fixed by our choice of prior. We will here choose the priors on all of the $\gamma$ parameters to be flat everywhere inside a fixed region that will be chosen by requiring that the PN approximation remain valid in the modified theory, i.e.~requiring that the 1PN term is not much much larger than the leading PN order term.  To ensure this, we set the prior to zero when the following condition is violated
\begin{align}\label{eq:prior_cond}
    |\zeta_{\EdGB} g_{\EdGB}(\boldsymbol{\theta}_{\GR}) (\pi \mathcal{M} f_{\text{lim}} )^{-7/3}| > |\delta \phi_{0\PN} (\pi \mathcal{M} f_{\text{lim}})^{-5/3}|
    \nonumber \\
\end{align}
where $f_{\text{lim}}$ is an approximate upper limit on the regime of validity for the expansion, and $\delta \phi_{0\PN}$ is the term proportional to $u^{-5}$ shown in Eqs.~\eqref{eq:GHOv1},~\eqref{eq:GHOv2}, and~\eqref{eq:GHOv3} for each parametrization, respectively.
This translates directly to the following criteria,
\begin{subequations}
\begin{align}\label{eq:prior_cond_explicit}
    |\gamma_{\text{v1}}| &< |g_{\EdGB}(\boldsymbol{\theta}_{\GR}) (\pi \mathcal{M} f_{\text{limit}})^{-2/3}|\,, \\
    % \nonumber
    |\gamma_{\text{v2}}| &< |(\pi \mathcal{M} f_{\text{limit}})^{-2/3}|\,,\\
    % \nonumber
    |\gamma_{\text{v3}}| &< |(\phi_{1\PN})^{-1} (\pi \mathcal{M} f_{\text{limit}})^{-2/3}| \,.
\end{align}
\end{subequations}
To keep the analysis simpler, we evaluate the conditions in Eq.~\eqref{eq:prior_cond} at the best fit values for the source parameters coming from a separate parameter estimation analysis assuming GR.
In this work, the frequency $f_{\text{lim}}$  will be set to the value of the GW frequency emitted by the binary at a separation of $50 M_{\odot}$. 
For each version of our parametrization, this corresponds to the allowed ranges $\gamma_{\text{v}i}$ of $[-1.31,1.31]$, $[-90.67,90.67]$, and $[-14.40,14.40]$, respectively.
Because each implementation uses a different definition of $\gamma_{{\rm v}i}$, a flat prior on $\gamma_{{\rm v}i}$ will not have the same impact on $\sqrt{\alpha_{\EdGB}}$ and will result in slightly different constraints once the posterior has been marginalized over $\gamma$.
This can be seen in Fig.~\ref{fig:GHO_priors}, which shows a direct comparison of the mapping of a uniform prior from each parametrization to each of the other parametrizations.

\begin{figure}
\includegraphics[width=\linewidth]{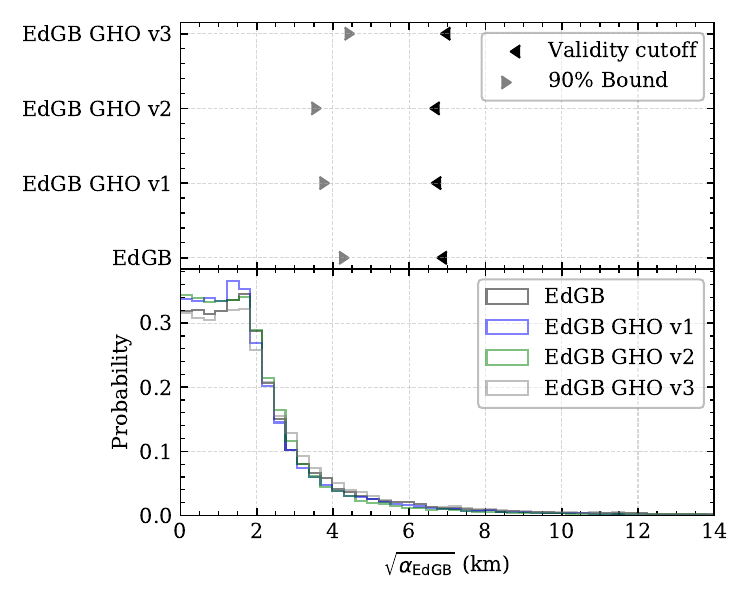}
\caption{
    The results of analyzing the GW151226 event with EdGB and three different versions of EdGB with some generic, higher PN order modification.
    The histograms in the lower panel show the posterior probability on the value of the EdGB coupling constant, $\sqrt{\alpha_{\EdGB}}$ in km.
    All three iterations include the -1PN effect, and the three versions labeled GHO (generic higher order) incorporate some generic modification at 0PN (relative to the Newtonian term in GR), outlined in the text.
    The upper panel shows the 90\% confidence constraint on $\sqrt{\alpha_{\EdGB}}$, and the maximum value the coupling constant can take and still satisfy the small coupling approximation.
    We see good agreement between these four methods, giving us confidence that our results in this work will be robust to any additional modifications that are derived in the future.
    }
    \label{fig:GHO_comp}
\end{figure}

The inferred value of  $\sqrt{\alpha_{\EdGB}}$ for these three different parametrization schemes applied to the GW151226 event are shown in Fig.~\ref{fig:GHO_comp}. 
As is clearly shown in the figure, the results are indeed robust to higher PN order corrections to the waveform, at least to next-to-leading order. The physical reason for this is that the EdGB correction enters at -1PN order, so the low frequency part of the signal is what is dominating the constraint. In this low frequency regime, the higher PN order terms are indeed subdominant, and do not affect the constraint on the leading PN order term. Indeed, this has also been observed in GR parameter estimation, where the chirp mass (which enters at leading 0PN order) is fixed by the low-frequency part of the signal, with the symmetric mass ratio (which enters at 1PN order) mattering only at higher frequencies. 

\begin{figure}[t]
    \centering
    \includegraphics[width=\columnwidth]{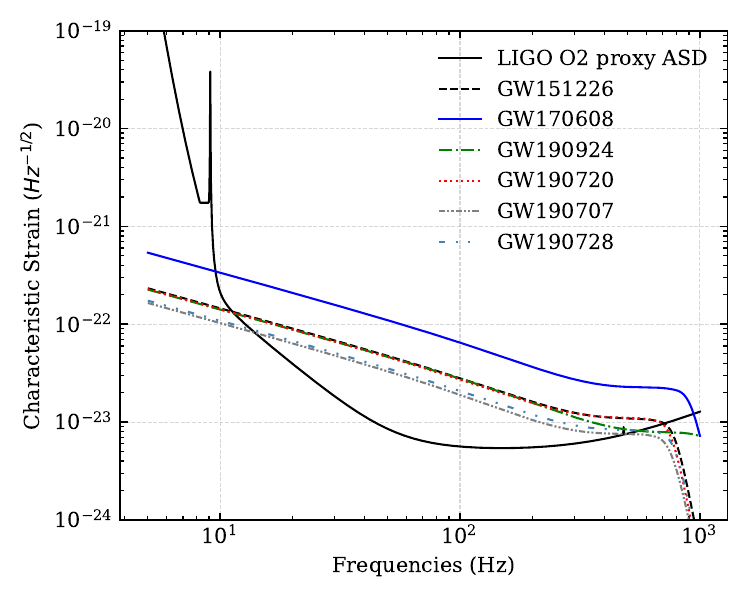}
    \caption{
        The characteristic strain compared against an example sensitivity curve representative of the LIGO detectors in O2. 
        The contribution to the SNR from the merger is negligible for almost all the events, ensuring that our analysis is robust to our lack of knowledge of the highly nonlinear dynamics of the merger itself.
        }
    \label{fig:character_strain}
\end{figure}

One other concern that could be raised is the lack of knowledge about the highly nonlinear regime of the merger of BBH in either of these theories of gravity.
This part of the coalescence of BBH cannot be described perturbatively, but instead requires numerical simulations.
In this work, we targeted the lowest mass sources from the events catalog for the reasons outlined in Sec.~\ref{sec:single_events}.
Because of this, the merger typically falls outside the range of sensitivity for the current 2G network of detectors, and would not impact this study. 
It is true that future work, performed with higher SNR sources, would need to be more cognizant of these issues, but the current restrictions due to detector limitations has ensured this work is robust to these uncertainties.
To illustrate our reasoning, we have shown in Fig.~\ref{fig:character_strain} the characteristic strain, given by $h_c = \sqrt{f} \sqrt{|h_{+}|^2 +|h_{\times}|^2}$, against a proxy noise curve for the LIGO detectors in the second observation run. 
This can be informative as the contribution to the SNR as a function of frequency for a source can be approximately seen as the ratio of the two curves.
The majority of the SNR for the sources used in this work comes from the early inspiral, with many of the sources' actual mergers contributing very little. 

%%%%%%%%%%%%%%%%%%%%%%%%%%%%%%%%%%%%%%%%%%%%%%%%%%%%%%
\section{Conclusions and Future Directions}\label{sec:conclusions}
%%%%%%%%%%%%%%%%%%%%%%%%%%%%%%%%%%%%%%%%%%%%%%%%%%%%%%

In this work, we have conducted a full, numerical exploration of the posterior surface of the coupling constants in two viable extensions to GR (dCS and EdGB) for six GW sources from GWTC-1~\cite{LIGOScientific:2018mvr} and GWTC-2~\cite{Abbott:2020niy}, utilizing the natural parameter-basis adapted to the problem. 
This custom analysis reduces the errors introduced in the recycling of results from generic analysis and provides the most robust and reliable results to date.
To further verify the conclusions of this work, we investigated the impact of changing the base (GR) waveform (onto which our modifications were appended), as well as the impact of our lack of knowledge of higher PN order corrections that would be introduced to the waveform as the result of modifications to the field equations.

We inferred to $90\%$ confidence that the square root of the coupling constant for EdGB is less than $1.7$ km, i.e.~$\sqrt{\alpha_{\EdGB}} < 1.7$ km.
Our work on EdGB gravity has produced a constraint on the coupling constant that is now the most stringent to date, as well as less prone to systematic error than previous analysis~\cite{Yagi:2012gp}.
Unfortunately, dCS gravity has continued to evade our efforts to constrain its coupling constant through the exclusive use of GW observations.
The constraints we place here, along with those of past works, have reliably restricted the length scale of the theory to roughly the size of the horizon of astrophysical BHs. 
Future work might improve on this bound through the use of merger-ringdown measurements~\cite{Carson:2020ter} (for which, more work will be needed in perturbation theory~\cite{Molina:2010fb,Blazquez-Salcedo:2016enn,Cano:2020cao,Wagle:2021tam,Pierini:2021jxd} and numerical simulations~\cite{Witek:2018dmd,Okounkova:2019dfo,Okounkova:2019zjf,Okounkova:2020rqw,East:2020hgw,Silva:2020omi}, but continued use of inspiral measurements will only serve to slowly push the limit further and further down in small increments.
From the point of view of inspiral only measurements, the theory is now at the mercy of the statistics of large catalogs.

Our results open the door for many future directions of research.
Of course, we are still hoping to constrain dCS gravity purely through GWs, but we must continue to wait for more favorable sources, such as those pointed out in~\cite{Alexander:2017jmt}.
Of particular interest would be a NSBH source with a reasonable SNR, as the unequal footing of NS and BH in dCS gravity works to the benefit of our efforts to constrain the theory.
With sufficient SNR, a NSBH event would provide the best avenue for constraining dCS gravity with a single source.

\noindent {\bf{\emph{Note added after completion:}}} Right before we submitted this paper, Ref.~\cite{Wang:2021yll} appeared in the arXiv. That work is different from the one presented here in various respects. First, Ref.~\cite{Wang:2021yll} considered single events, instead of stacking as we do in this paper. Second, Ref.~\cite{Wang:2021yll} did not consider the possible effect of waveform systematics on their constraints; neither due to higher PN order corrections in the non-GR part of the waveform, nor due to different models for the GR part of the waveform. We studied this in great detail.
Finally, and perhaps most importantly, Ref.~\cite{Wang:2021yll} used two particular events in the GWTC-2 catalog, GW190814 and GW190425, both of which we purposely excluded from our analysis (see Sec.~\ref{sec:source_selection} and Appendix~\ref{app:source_classification}). 

We excluded GW190814 for two reasons: the uncertainties in the GR part of the waveform greatly affected non-GR constraints in our analysis and the uncertain nature of the lighter object.  
In fact, the former reason affected the tests of GR that the LVC carried out with GWTC-2, which is why the LVC made explicit mention of this event in their analysis (see Appendix C and, in particular,~Fig.~19 of~\cite{Abbott:2020jks}). 
We excluded GW190425 because this event was very low-mass and, in fact, it was identified as a binary NS inspiral by the LVC (at $>99\%$ confidence at the GraceDB website~\cite{GraceDB}). 
As is well-known from~\cite{Yagi:2015oca}, NSs do not possess a monopole scalar charge in the version of EdGB gravity we considered in this work.
Therefore, NS binaries in this theory do not introduce scalar dipole modifications to the 
GW phase. 
As a consequence, the only way the analysis of Ref.~\cite{Wang:2021yll} is correct is if one assumes \textit{a priori} that the binary was composed of BHs, which even with the analysis of Ref.~\cite{Han:2020qmn} seems like a very strong assumption. 

%%%%%%%%%%%%%%%%%%%%%%%%%%%%%%%%%%%%%%%%%%%%%%%%%%%%%%
\acknowledgements
This work made use of the Illinois Campus Cluster, a computing resource that is operated by the Illinois Campus Cluster Program (ICCP) in conjunction with the National Center for Supercomputing Applications (NCSA) and which is supported by funds from the University of Illinois at Urbana-Champaign.
We acknowledge financial support from  NSF Grant PHY-1759615 and PHY-1949838, NASA Grants 80NSSC18K1352, 
NNX16AB98G and 80NSSC17M0041 and NASA ATP Grant No.~17-ATP17-0225.
\appendix

\section{Concerns over source classification}\label{app:source_classification}

Unique GW events can offer new and exciting avenues to test fundamental physics, but they do not come without their own complications.
For example, binaries near the mass gap, where one component might be a heavy NS or a light BH, can potentially provide a strong test of GR.
However, such tests would undoubtedly be hindered if one were unsure of the classification of each component object (BH or NS).
In many cases, one can be confident in the exact nature of the binary, for example in the case of an electromagnetic counterpart or a robust detection of the tidal deformability of one of the objects.
Unfortunately, two intriguing events we considered do not fall in this category, and we opted to not include them in the main body of this work, as explained in Sec.~\ref{sec:source_selection}.
In this Appendix, we will discuss some implications of these complications.

Let us first discuss GW190814, which was found to be a highly asymmetric binary. 
The uncertain nature of the lighter compact object in the binary poses some complications to our analysis.
This is because a NS and a BH have different properties in dCS and EdGB theories. 
In particular, BHs in EdGB gravity support monopole scalar charges (see e.g.~\cite{Kanti:1995vq,Yunes:2011we,Prabhu:2018aun}), while NSs do not~\cite{Yagi:2015oca,Saffer:2019hqn}. 
This means that if one knew that the binary was mixed, one could use NSBH waveforms with EdGB modifications, setting the sensitivity $s_2^{\EdGB}$ in Eq.~\eqref{eq:EdGB_beta} to zero.
On the other hand, if one assumes that the binary was composed of two BHs, then one must use a BBH waveform with EdGB modifications that allow both sensitivities $s_1^{\EdGB}$ and $s_2^{\EdGB}$ in Eq.~\eqref{eq:EdGB_beta} to be non-zero.
Therefore, our lack of knowledge about the exact composition of the lighter object can significantly affect any parameter
estimation in EdGB, hence casting doubt over any constraint placed on the theory from such systems.

\begin{figure}[h]
    \centering
    \includegraphics[width=\linewidth]{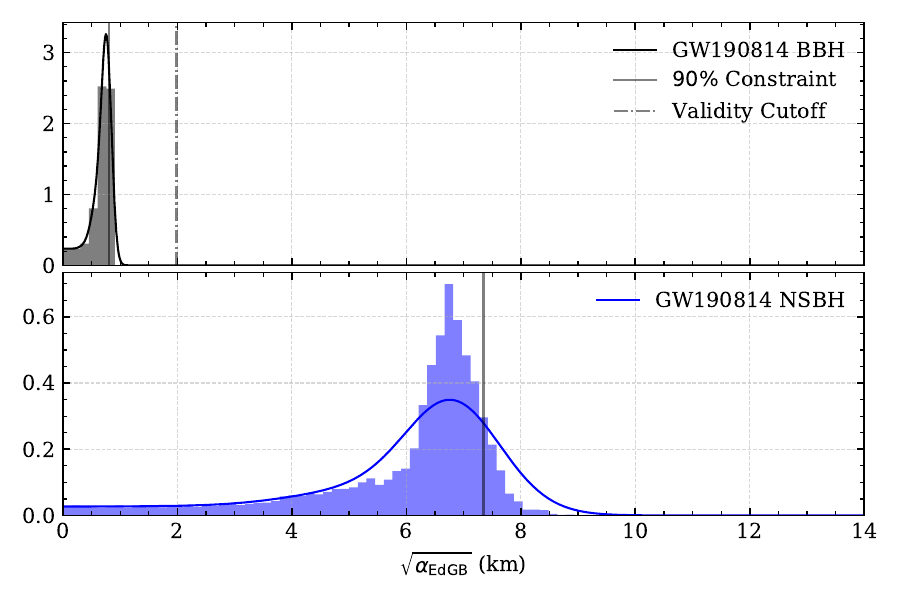}
    \caption{
    The posterior distributions on $\sqrt{\alpha_{\EdGB}}$ for the GW190814 event.
    The top panel was derived from an analysis that assumed GW190814 was a BBH system, while the lower panel shows results that were derived assuming the source is a NSBH system.
    The small coupling criteria for this source, when considering it as a NSBH source, is approximately 18 km, and lies beyond the edge of the plot.
    As is immediately apparent, the upper limit on $\sqrt{\alpha_{\EdGB}}$ based on each of these assumptions differs by an order of magnitude.
    }
    \label{fig:GW190814_comp}
\end{figure}

As we mentioned above, for the GW190814 event there are two reasonable possibilities: that the source is a BBH or a NSBH binary.
There are various reasons to believe either one over the other 
(see e.g.~\cite{Fishbach:2020ryj,Most:2020bba,Broadhurst:2020cvm,Abbott:2020khf}), 
and as such, any analysis making conclusions based on the \emph{a priori} assumption of one classification over the other should carefully consider both possibilities.
To show how the assumptions on the binary constituents can dramatically affect the constraints on EdGB, we analyzed GW190814 for \textit{both BBH and mixed binary scenarios.}
The posteriors on $\sqrt{\alpha_{\EdGB}}$ are shown in Fig.~\ref{fig:GW190814_comp}.
We can immediately see that the prior assumption of the source composition seriously impacts our constraints, with over an order of magnitude difference between the two bounds.
This fact casts doubt over any constraint on this particular theory using this event.

Moreover, possible waveform systematics can clearly be seen in our posteriors.
The LVC analysis of GW190814 led to posteriors on a $-1$PN order modification to GR that excluded GR to $90\%$ confidence~\cite{Abbott:2020jks}. 
We performed a similar Bayesian analysis on GW190814 both with a generic $-1$PN order modification and with an EdGB modification. 
5
In both cases we find that the posteriors on the non-GR parameter excludes GR, in agreement with the LVC analysis. 
This result is not believed to be a real deviation from GR, but rather a consequence of waveform systematics and covariances between parameters. 
While one could ignore these deviations and calculate upper limits on $\sqrt{\alpha_{\EdGB}}$ regardless, the obvious impact of waveform systematics is of note~\cite{Moore:2021eok}.
This stark difference in the two posteriors along with the issues of waveform systematics, in a context where neither interpretation of the event GW190814 is unquestionable, led us to omit this source from the main body of this work.

In the case of GW190425~\cite{Abbott:2020uma}, the situation becomes even more stark. 
While this event might lead to stringent constraints on one or both of these theories if it were considered a NSBH binary, there is the strong possibility that this source is actually a NSNS binary.
In the latter case, one cannot meaningfully constrain either EdGB or dCS.
This is an even more extreme case where the difference between the constraints when using one assumption over the other is infinite.
Considering the significant impact of the prior assumption on the composition of the binary on final constraint conclusions, and the current evidence suggesting this is actually a NSNS binary~\cite{Abbott:2020uma}, we neglected this source as well.

\section{Methods for combining posteriors}\label{app:stacking_comparison}

\begin{figure}
    \centering
    \includegraphics[width=\columnwidth]{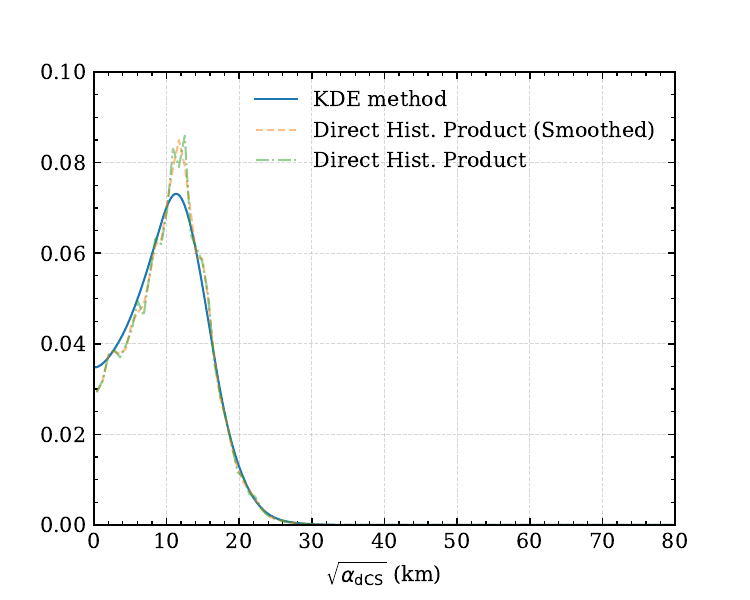}
    \caption{
    Comparison of the two different methods for combining posterior distributions from independent experiments into a single, joint distribution on $\sqrt{\alpha_{\dCS}}$.
    The two methods in question involve the fitting of functions to the individual distributions (KDE method) and the direct product of the histograms (Direct Hist.~Product), as well as a smoothed version of the latter used for the calculation of the confidence intervals.
    While there seems to be a bias towards higher values of $\alpha_{\dCS}$, any conclusion from this distribution should be tempered, as this analysis still doesn't provide reliable results, due to violations to the small coupling approximation.
    }
    \label{fig:joint_comp_dCS}
\end{figure}
\begin{figure}
    \centering
    \includegraphics[width=\columnwidth]{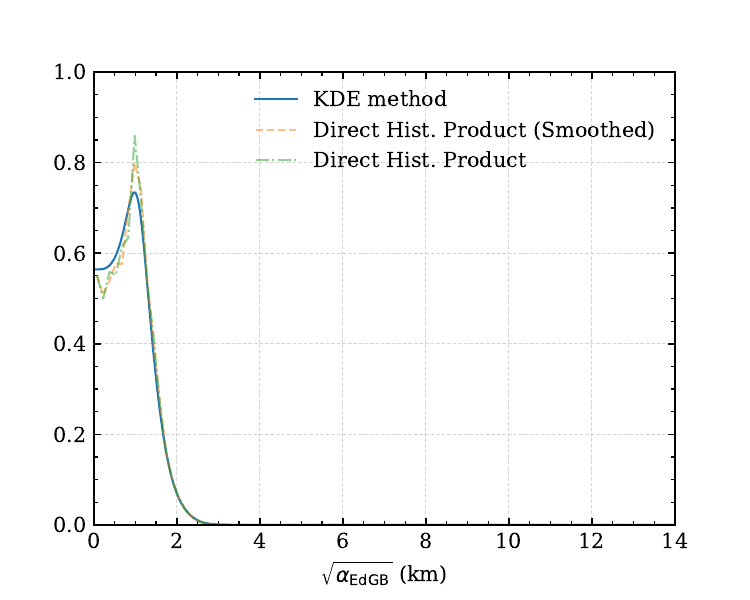}
    \caption{
    Comparison of the two different methods for combining posterior distributions from independent experiments into a single, joint distribution on $\sqrt{\alpha_{\EdGB}}$.
    The two methods in question involve the fitting of functions to the individual distributions (KDE method) and the direct product of the histograms (Direct Hist. Product), as well as a smoothed version of the latter used for the calculation of the confidence intervals.
    }
    \label{fig:joint_comp_EdGB}
\end{figure}

To combine posteriors on a shared parameter from multiple observations, there are two commonly employed methods of calculating the joint posterior, as outlined in Sec.~\ref{sec:multiple_events}.
One involves the fitting of some ansantz function or KDE approximation to the discrete, marginalized likelihoods, then multiplying the analytic fits together for the cumulative constraint.
Meanwhile, the other method involves the direct multiplication of the discrete, marginalized likelihoods together, then working with the final histogram directly.

Both have their benefits and drawbacks, and we present both in Fig.~\ref{fig:joint_comp_dCS} and Fig.~\ref{fig:joint_comp_EdGB} for dCS and EdGB, respectively.
For the former method, we have used a KDE to estimate the marginalized likelihoods for each individual event, and for the latter method, we smoothed the \emph{final} histogram with a Savitzky-Golay filter with a polynomial order of 5 and a window of 7 before calculating confidence intervals for numerical stability.

As noted in the main body of this paper, one drawback to the KDE method is the lack of ability to account for hard cutoffs, for example the boundary at $\sqrt{\alpha} = 0$.
The KDE approach led to a joint posterior that rapidly dropped to zero for small $\sqrt{\alpha}$, and to remedy this, we fit the KDE to a distribution described by the set of samples twice as large as the original set coming from the MCMC.
We used the distribution $\{s_i\} \cup -\{s_i\}$, which amounts to reflecting the distribution across 0.
We then renormalized the distribution numerically in the range $\sqrt{\alpha} \in [0,\infty]$.
This effectively solved the boundary issue in this particular use case, as the KDE approximation smoothly approached a final value at $\sqrt{\alpha}=0$.

The final result from the two methods agreed well for the six sources considered in this work.
To be maximally cautious, we only use the worse of the two methods when quoting $90\%$ confidence constraints on EdGB.

%%%%%%%%%%%%%%%%%%%%%%%%%%%%%%%%%%%%%%%%%%%%%%%%%%%%%%
\bibliography{refs}
\end{document}